\documentclass[%
 preprint, 
 amsmath,amssymb,
 aps, physrev,
]{revtex4-2}

\usepackage{graphicx}
\usepackage{dcolumn}
\usepackage{bm}

\usepackage[english]{babel}
\usepackage{siunitx}
\usepackage{longtable,tabularx}
\usepackage{upgreek}
\usepackage{IEEEtrantools}
\usepackage{xcolor}
\usepackage[colorlinks=true,citecolor=blue,urlcolor=black]{hyperref}
\newcommand{\ud}{\mathrm{d}}
\newcommand{\ue}{\mathrm{e}}
\newcommand{\ui}{\mathrm{i}}
\newcommand{\uexp}{\mathrm{exp}}
\newcommand{\uph}{(h/\beta)}

\newcommand{\Q}{Q}
\newcommand{\D}{d}
\newcommand{\<}{\negmedspace{}}
\begin{document}

\preprint{Preprint}

\title{\textbf{Three-dimensional modelling of serrated trailing-edge noise based on the Wiener-Hopf technique} 
}%

\author{Sicheng Zhang}
\author{Benshuai Lyu}%
 \email{Corresponding author: b.lyu@pku.edu.cn}
\affiliation{%
 State Key Laboratory of Turbulence and Complex Systems, School of Mechanics and Engineering Science, Peking University, Beijing 100871, China
}%

\date{\today}

\begin{abstract} 
In this paper, a semi-analytical model based on the Wiener-Hopf technique is developed to predict the turbulent boundary layer trailing edge noise from serrated edges, aiming to account for the correct three-dimensional noise source and propagation effects. The scattered surface pressure over a semi-infinite flat plate is first obtained using the Green's function developed for the acoustic scattering by a serrated edge. A radiation integral over the flat plate of a finite size is subsequently performed to obtain the far-field noise using Amiet's approach, capturing the correct three-dimensional source and propagation effects. The model is subsequently validated by comparing it against the two-dimensional Wiener-Hopf-based model under various serration sizes and frequencies. Far-field spectral predictions show close agreement between the three- and two-dimensional models at moderate observer distances around $r/c=1$,  where $r$ and $c$ represent the observer distance and the chord length, respectively. However, unlike the two-dimensional model, the present model successfully captures the far-field $1/r$ decay in noise amplitudes. In addition, the predicted directivity agrees well with the two-dimensional model at most observer angles, but also captures the correct dipolar behaviour at upstream angles and additional high-frequency lobes due to interference patterns induced by the finite flat plate. Compared to the previous three-dimensional serrated models, the present model is based on the Wiener-Hopf technique and achieves a speed-up ratio of two orders of magnitude. It is hoped that such a model may be used to enable an efficient numerical optimisation of the serration shape in realistic applications. 
\end{abstract}

\maketitle
\clearpage

\section{Introduction}
Trailing-edge (TE) noise refers to the noise generated when turbulent boundary layers convect past the trailing edge of an aerofoil \citep{howe_review_1978}. It represents an important noise source in many modern industrial applications, such as wind turbines. As the blade of the wind turbine becomes larger, the tip of the blade rotates at a larger velocity. TE noise increases quickly with this velocity and becomes the dominant noise source for modern wind turbines~\citep{oerlemans_reduction_2009}.

Developing analytical models that accurately predict TE noise is essential for a
comprehensive understanding of TE noise and lays the foundation for developing
techniques to reduce wind turbine noise. Pioneering work by Amiet modelled TE
noise scattered from a flat plate with a straight trailing edge.
\citet{amiet_noise_1976} treated the aerofoil
as a semi-infinite flat plate to calculate the scattered surface
pressure using the Schwarzschlild technique. Far-field noise is obtained by
performing a radiation integral over a flat plate of a finite size. This model
established an analytical transfer function mapping the wavenumber spectral
density of the wall pressure fluctuations beneath the turbulent boundary layer
to the far-field noise spectra. The model was further developed by incorporating
the effect of the incident pressure fluctuation\citep{amiet_effect_1978}. To
account for the effect of a finite aerofoil in calculating the scattered surface
pressure, \citet{roger_back-scattering_2005} extended Amiet's model by
incorporating the back-scattered pressure from the leading edge. Their results
revealed that the back-scattered pressure alters the far-field sound only at
very low frequencies and is negligible when $kc>1$, where $kc$ is the acoustic
Helmholtz number based on the chord length of the finite flat plate.

Extensive research has been conducted to reduce TE noise. Serrations, inspired
by the silent flight of owls~\citep{jaworski_aeroacoustics_2020}, are widely
used as a passive approach. Numerous experiments
\citep{dassen_results_1996,arce_leon_flow_2016,oerlemans_reduction_2009,gruber_airfoil_2010,gruber_airfoil_2012,chong_airfoil_2013,leon_effect_2017,moreau_noise-reduction_2013,chong_aeroacoustic_2015,avallone_benefits_2017,ZHOU2025110851}
have shown that the serrated designs implemented at the trailing edge can reduce
TE noise effectively. \citet{dassen_results_1996} measured the noise reduction
effects of serrations on both aerofoils and flat plates, noting reductions of up
to 8 dB and 10 dB, respectively. Conducting a similar experiment,
\citet{arce_leon_flow_2016} revealed that sawtooth trailing-edge serrations on a
NACA 0018 aerofoil reduce broadband noise by up to 7 dB at low and mid
frequencies. However, in the high-frequency regime, noise increased when
serrations were misaligned with the flow direction. Experiments by
\citet{oerlemans_reduction_2009} examined the sound field of a full-scale wind
turbine with standard, optimised, and serrated blades. An average noise
reduction of 3.2 dB was observed through the incorporation of serrations on the
trailing edge of blades.

In recent years, various experiments have been conducted to identify optimal
trailing-edge serration geometries, such as iron-shaped serrations
\citep{avallone_benefits_2017} and ogee serrations \citep{ZHOU2025110851}.
\citet{avallone_benefits_2017} reported that iron-shaped serrations achieve
approximately 2 dB noise reduction in the low- and mid-frequency ranges. This
improvement was attributed to weaker root-flow interactions and a lower
noise-source intensity near the serration root. \citet{ZHOU2025110851} examined
ogee serrations applied to NACA 0012 aerofoils in a wind tunnel and observed a
1–3 dB noise reduction, outperforming sawtooth serrations when the edges were
sharp.
    
In addition to experiments, numerical simulations have been used to study noise
reduction using
serrations~\citep{jones_acoustic_2012,sanjose_direct_2014,avallone_noise_2018}.
\citet{jones_acoustic_2012} conducted Direct Numerical Simulations (DNS) of the
flow around a NACA0012 aerofoil to investigate the impact of serrations on TE
noise. They showed that the turbulent boundary layer statistics were nearly
unaffected, and the noise was effectively reduced by the serrations.
\citet{sanjose_direct_2014} performed a DNS on a serrated controlled diffusion
using a Lattice-Boltzmann method and reported a similar conclusion.
\citet{avallone_noise_2018} also performed a compressible simulation using the
Lattice-Boltzmann method to investigate the noise reduction from different types
of sawtooth serration. The combed-sawtooth serrations were found to reduce more
noise than the conventional sawtooth serrations at low- and mid-frequencies. 
	
Experimental and numerical findings consistently demonstrate efficient reduction
of TE noise through serrations. However, analytical prediction models are often
needed in order to design the optimal serration geometry.
\citet{howe_noise_1991,howe_aerodynamic_1991} developed an early analytical
model to predict serrated TE noise, revealing that sharp sawtooth serrations
were more effective in suppressing TE noise. However, Howe's model significantly
overpredicted the noise reductions. Based on Amiet's approach,
\citet{lyu_prediction_2016} developed a serrated TE noise model, significantly
improving prediction accuracy through an iterative procedure using the
Schwarzschild technique and Fourier transforms. It was shown that the noise
reduction arises from the destructive interference of the scattered pressure in
the vicinity of the serrated edge. The iterative procedure involved the
evaluation of nested summations, incurring substantial computation time for
serration shape optimisation. \citet{tian2022theoretical,tian_prediction_2022}
extended Lyu's model to predict the noise emission from rotating blades with
serrated trailing edges. The model predicted the noise directivity of rotating
blades and the contribution of the Doppler effect on the noise
directivity.

Recently, \citet{ayton_analytic_2018} developed a model to evaluate serrated TE
noise using the Wiener-Hopf method. The Wiener-Hopf technique enabled the assessment of far-field sound with two infinite sums and one infinite integral. \citet{lyu_rapid_2020} further extended Ayton's model by explicitly evaluating the infinite integral and one of the infinite sums. This simplified model is much more efficient in TE noise spectra prediction. However, due to the semi-infinite flat plate assumption used in ~\citet{ayton_analytic_2018}, both the original and simplified models are strictly two-dimensional. This leads to far-field sound scaling as $1/\sqrt{r}$ instead of $1/r$ when $r$ approaches $\infty$, where $r$ denotes the observer distance. In many applications, such as rotating fans or propellers, the distance and relative angle between the observer and the blade constantly change. To correctly predict noise spectra or directivity in these applications, accounting for the correct three-dimensional noise source and propagation effects is crucial. To predict rotorcraft noise, \citet{li_extensions_2022} extended the two-dimensional model of \citet{lyu_rapid_2020} and developed a heuristic three-dimensional model in a heuristic manner. However, to improve accuracy and robustness, an analytically or semi-analytically rigorous approach is desired.

	It is well-established that the three-dimensional effects are accounted for (at
	least partially) in Amiet's model\citep{amiet_noise_1976}. This is because the scattered pressure on a
	semi-infinite plate is evaluated first, and the far-field sound is obtained by
	using a finite surface integral. To do this, the near-field surface pressure is
	needed. Such a capability is not available in \citet{ayton_analytic_2018} or 
	\citet{lyu_rapid_2020}, because the steepest descent method is used to evaluate
	the inverse Fourier transform, which is only valid in the far field.
    To overcome the difficulty, \citet{lyu_analytical_2023} developed a Green's function using the Wiener-Hopf technique. This Green's function can be used to evaluate the near-field pressure, thereby enabling the development of a three-dimensional model analytically. The Wiener-Hopf method-based model is potentially much faster, rendering the long-sought numerical optimisation of serration geometry possible.  
	
	In this paper, we aim to bridge the gap and use the recently developed Green's function~\citep{lyu_analytical_2023} to develop a semi-analytical three-dimensional model for serrated trailing-edge noise predictions. This paper is organised as follows: Section \ref{sec:analytical derivation} presents the mathematical model and its derivation. In Section \ref{sec:Model Validation}, a comparison between the model predictions and the two-dimensional Wiener-Hopf method-based  model~\citep{ayton_analytic_2018,lyu_rapid_2020} is provided. The computational efficiency of this model is also discussed. The final section presents a brief conclusion and outlines potential future work.

\section{Analytical derivation}\label{sec:analytical derivation}
\subsection{Mathematical model}

	\begin{figure}
		\centerline{\includegraphics[width=0.5\textwidth]{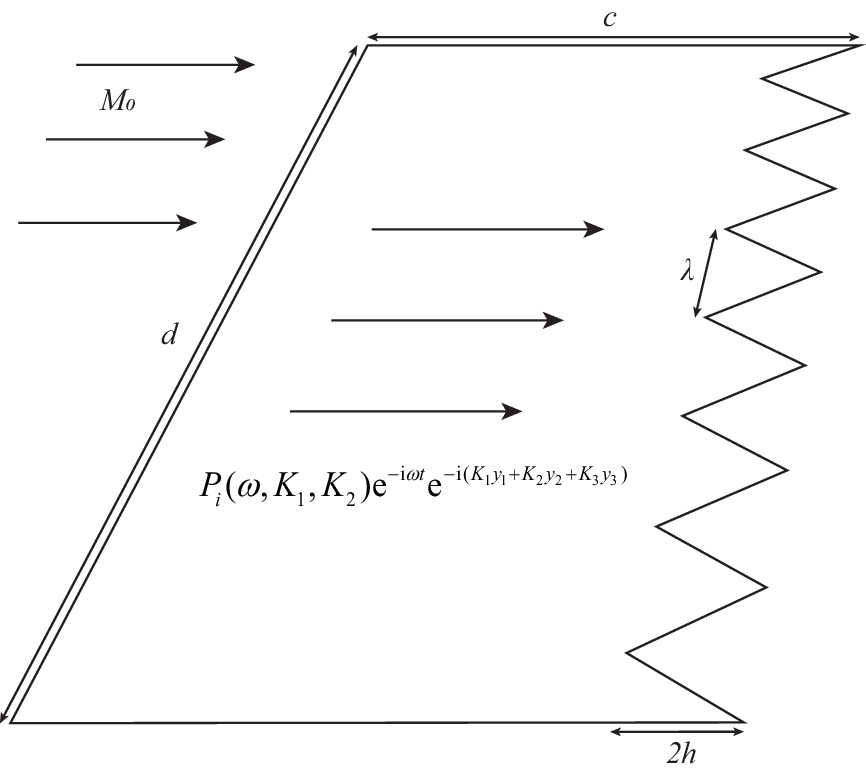}}
		\caption{The simplified model of a flat plate with sawtooth serrations positioned in a uniform streamwise flow. The chord and span lengths of the plate are $c$ and $\D$, respectively, while the wavelength and root-to-tip amplitude of the serration are denoted by $\lambda$ and $2h$, respectively. The coordinates $y_1, y_2, y_3$ are aligned with the streamwise, spanwise, and normal-to-plate directions, respectively. The far-field observer is located at $(x_1, x_2, x_3)$.}
		\label{fig:model}
	\end{figure}
	We consider a flat plate with a serrated trailing edge in a uniform flow, as illustrated in figure \ref{fig:model}. The uniform velocity is non-dimensionalised by the speed of sound $c_0$, and the resulting Mach number is $M_0$. The plate has a chord length of $c$ and span length of $\D$, while the sawtooth serrations exhibit a wavelength of $\lambda$ and a root-to-tip amplitude of $2h$. The streamwise, spanwise, and normal-to-plate coordinates are represented by $y_1$, $y_2$ and $y_3$, respectively. The far-field observer is located at ($x_1$, $x_2$, $x_3$).

    Following the definition in~\citet{lyu_analytical_2023}, we assume that the incident wall pressure gust is given by
	\begin{equation}
		P_{in}=P_i(\omega,K_1,K_2)\ue^{-\ui\omega t}\ue^{-\ui\left(K_1y_1+K_2y_2+K_3y_3\right)},
		\label{eqn:P_i}
	\end{equation}
	at $y_3=0$. Here, $\omega$ is the angular frequency, and ($K_1$, $K_2$, $K_3$) represent the streamwise, spanwise, and normal-to-plate wavenumbers, respectively. It should be noted that, the incident gust is convected by the mean flow. Hence, in what follows, the streamwise wavenumber $K_1$ is decomposed into $k_1/\beta+kM_0/\beta^2$, where $\beta^2=1-M_0^2$ and $k_1$ is the transformed wavenumber as in \citet{lyu_analytical_2023}.

	In the rest of this paper, unless noted otherwise, all variables are non-dimensionalised using the serration wavelength $\lambda$, speed of sound $c_0$, and fluid density $\rho_0$. First, because of linearity, we assume $P_i$ to be 1 and obtain the transfer function from the gust to the far-field sound pressure. Subsequently, the wavenumber-frequency spectrum of the boundary-layer pressure fluctuations is substituted into the transfer function as the input to obtain the far-field sound spectrum from the serrated trailing edge.

	\subsection{The scattered pressure on the surface}
	\label{section: The scattered pressure on the surface}
\begin{figure}
		\centerline{\includegraphics[width=0.6\textwidth]{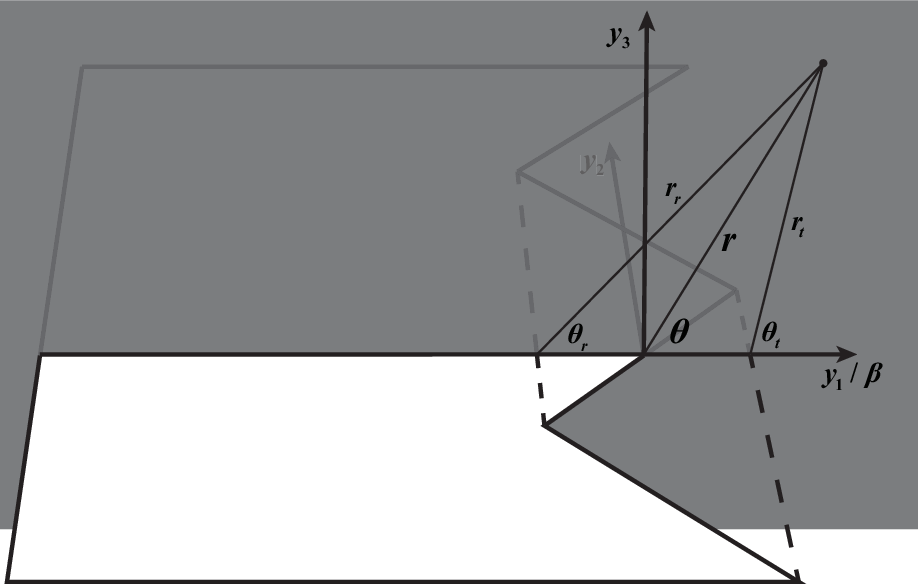}}
		\caption{The polar coordinates used in the derivation. $(r,\theta)$ is the polar coordinate frame in the stretch $y_1/\beta-y_3$ plane. Two auxiliary polar coordinate frames originate from $(h/\beta,0)$ and $(-h/\beta,0)$, respectively. }
		\label{fig:model_2}
	\end{figure}

	In this section, we assume $P_i=1$ and calculate the scattered pressure on a semi-infinite flat plate with a serrated trailing edge, as shown in figure~\ref{fig:model_2}. Using the Green's function developed to approximate solutions at high frequencies, one can write the scattered pressure as
	\begin{equation}
		\begin{aligned}
			G_s(r,\theta,y) &= \dfrac{-\ui}{2\uppi}\ue^{-\ui kM_0y_1/\beta^2}\sum_{n=-\infty}^{\infty}\ue^{\ui \chi_n y_2}\int_{-\infty}^{\infty}\dfrac{E_n(s)}{s-k_1}\dfrac{\sqrt{k_1-\kappa_n}}{\sqrt{s-\kappa_n}}\ue^{-\ui sy_1/\beta-\eta_n y_3}\ud s,
		\end{aligned}
		\label{eqn:G_s_define}
	\end{equation}
	where $s$ is a complex wavenumber defined in \citet{lyu_analytical_2023}, and
	\begin{IEEEeqnarray*}{rlrl}
	    k&=\omega/c_0, \quad &\Bar{k}=k/\beta,\quad 	\eta_n &= \sqrt{s^2-\kappa_n^2},\\
	\quad\chi_n &= 2n\uppi-K_2,&\quad\kappa_n &= \sqrt{\Bar{k}^2-\chi_n^2}.
	\end{IEEEeqnarray*}
	The functions $E_n(s)$ depend on the serration shape and are given by
	\begin{equation}
		E_n(s) = \int_{0}^{1}\ue^{\ui(s-k)\Bar{h}F(y_2)}\ue^{-\ui 2n\uppi y_2}\ud y_2,
	\end{equation}
	where $\Bar{h}$ is defined as $\Bar{h}=h/\beta$, and $F(y)$ represents the serration profile. When the serration is of the standard swatooth shape, i.e.
	\begin{equation}
		F(y_2)=\left\{
		\begin{aligned}
			&4y_2,\qquad\qquad\quad -\frac{1}{4}<y_2\leq\frac{1}{4},\\
			&-4y_2+2,\quad\qquad \frac{1}{4}<y_2\leq\frac{3}{4},
		\end{aligned}
		\right .
	\end{equation}
	the functions $E_n(s)$ can be evaluated as
	\begin{equation}
		E_n(s) = \dfrac{4(s-k_1)\Bar{h}\sin\left(\left(s-k_1\right)\Bar{h}-n\uppi/2\right)}{4(s-k_1)^2\Bar{h}^2-n^2\uppi^2}.
		\label{eqn:En}
	\end{equation}
	
	By analytically evaluating the integral in \eqref{eqn:G_s_define}, the Green's function $G_s$ for this sawtooth serration can be written as
	\begin{equation}
		\begin{aligned}
			G_s(r,\theta,y) &= \dfrac{1}{2\uppi}\ue^{-\ui kM_0y_1/\beta^2}\sum_{n=-\infty}^{\infty}-\ui(\sqrt{k_1-\kappa_n})\ue^{\ui \chi_n y_2}\\
			&\times \left[\uexp\left(-\ui\left(k_1 \Bar{h}+\dfrac{n\uppi}{2}\right)\right)H_n(r_t,\theta_t)-\uexp\left(\ui\left(k_1 \Bar{h}+\dfrac{n\uppi}{2}\right)\right)H_n(r_r,\theta_r)\right].
		\end{aligned}
		\label{eqn:G_s}
	\end{equation}
	Function $H_n(r,\theta)$ take different forms when $n\neq 0$ and $n=0$. For $n\neq 0$, $H_n(r,\theta)$ is a function defined by
	\begin{equation}
		H_n(r,\theta)=\dfrac{\ui}{\sqrt{2\kappa_n}n}\bigg(\dfrac{I(\kappa_n r,\theta;\Theta_n^+)}{\sin\frac{1}{2}\Theta_n^+}-\dfrac{I(\kappa_n r,\theta;\Theta_n^-)}{\sin\frac{1}{2}\Theta_n^-}\bigg),
		\label{eqn:H_n}
	\end{equation}
	where $\cos\Theta_n^\pm = (k_1 \pm n\uppi/2\Bar{h})/\kappa_n$, and $(r_t,\theta_t)$ and $(r_r,\theta_r)$ are two auxiliary polar coordinates defined in figure \ref{fig:model_2}. 
	Function $I(kr,\theta;\Theta)$ is the classical Fresnel solution denoting the pressure field scattered by a straight trailing edge \cite{noble_methods_1959}, which is defined as
	\begin{equation}
		I(kr,\theta;\Theta)=\dfrac{\ue^{-\ui\frac{\uppi}{4}}}{\sqrt{\uppi}}\bigg[\ue^{-\ui kr\cos(\Theta+\theta)}\Bar{F}(\sqrt{2kr}\cos\dfrac{\Theta+\theta}{2})-\ue^{-\ui kr\cos(\Theta-\theta)}\Bar{F}(\sqrt{2kr}\cos\dfrac{\Theta-\theta}{2})\bigg].
		\label{eqn:I_Fresnel}
	\end{equation}
	The Fresnel integral $\Bar{F}(x)$ takes the form of
	\begin{equation}
		\Bar{F}(x)=\int_x^\infty \ue^{\ui u^2}\ud u.
	\end{equation}
	Using change of variables, we can show that the error function $E(x)$ defined in \cite{lyu_prediction_2016} can be represented by $\bar{F}(x)$ as
	\begin{equation}
		E(x)=\int_0^x \dfrac{\ue^{\ui t}}{\sqrt{2\uppi t}}\ud t=\sqrt{\dfrac{2}{\uppi}}\left(\Bar{F}(0)-\Bar{F}(\sqrt{x})\right).
	\end{equation}
	The scattered pressure we are interested in is located on the plate surface. When the location is upstream of the serration root, both $\theta_r$ and $\theta_t$ take the value of $\uppi$. In contrast, when the location is downstream of the root, $\theta_t$ equals $\uppi$ but $\theta_r$ equals 0. Consequently, the function $H_n$ takes different forms within these two regions.
	When $\theta=0$, $I(kr,\theta;\Theta)$ vanishes, thus $H_n(r,0)$ vanishes. So we focus on the case with $\theta=\uppi$, in which case $I(kr, \pi, \Theta)$ reduces to
	\begin{IEEEeqnarray}{l}
			I(kr,\uppi;\Theta) = \dfrac{\ue^{-\ui\frac{\uppi}{4}}}{\sqrt{\uppi}}\ue^{\ui kr\cos\Theta}\cdot 2\int_0^{\sqrt{2kr\sin^2\frac{\Theta}{2}}} \ue^{\ui u^2}\ud u
			= (1-\ui)\ue^{\ui kr\cos\Theta}E(kr(1-\cos\Theta)).
		\label{eqn:I_n}
	\end{IEEEeqnarray}
	
	When the location is upstream of the serration root, substituting equation~\eqref{eqn:I_n} into equation~\eqref{eqn:H_n} leads to
	\begin{IEEEeqnarray}{rll}
			H_n(r,\uppi)&= \dfrac{\ui+1}{n}\ue^{\ui k_1r}
            \left[
            \dfrac{\ue^{\ui n\uppi r/2\Bar{h}}E(r(\kappa_n-k_1-n\uppi/2\Bar{h}))}{\sqrt{\kappa_n-k_1-n\uppi/2\Bar{h}}}-
			\dfrac{\ue^{-\ui n\uppi r/2\Bar{h}}E(r(\kappa_n-k_1+n\uppi/2\Bar{h}))}{\sqrt{\kappa_n-k_1+n\uppi/2\Bar{h}}}
            \right].\IEEEeqnarraynumspace
	\end{IEEEeqnarray}
	Therefore, the term in the square bracket in equation~\eqref{eqn:G_s} can be re-organised to be
	\begin{IEEEeqnarray}{rl}
			\dfrac{\ui(1-\ui)}{n}\dfrac{\ue^{-\ui\frac{y_1}{\beta}(k_1+n\uppi/2\Bar{h})}}{\sqrt{\kappa_n-k_1-n\uppi/2\Bar{h}}}&\bigg[ E(r_t(\kappa_n-k_1-n\uppi/2\Bar{h})) - E(r_r(\kappa_n-k_1-n\uppi/2\Bar{h}))\bigg] \IEEEnonumber\\
			- \ue^{\ui n\uppi}\dfrac{\ui(1-\ui)}{n}\dfrac{\ue^{-\ui\frac{y_1}{\beta}(k_1-n\uppi/2\Bar{h})}}{\sqrt{\kappa_n-k_1+n\uppi/2\Bar{h}}}&\bigg[ E(r_t(\kappa_n-k_1+n\uppi/2\Bar{h})) - E(r_r(\kappa_n-k_1+n\uppi/2\Bar{h}))\bigg] .\IEEEeqnarraynumspace
		\label{eqn:Q_n}
	\end{IEEEeqnarray}
	After substituting equation~\eqref{eqn:Q_n} into equation~\eqref{eqn:G_s}, we can get the final result of the $n$th ($n\neq 0$) term of $G_s$ as
	\begin{IEEEeqnarray}{rll}
			G_s^{(n)}(y) &=& \dfrac{1-\ui}{2n\uppi}\ue^{\ui\chi_n y_2}\ue^{-\ui kM_0y_1/\beta^2}\sqrt{k_1-\kappa_n}\times\IEEEnonumber\\
			&&\<\bigg[\dfrac{\ue^{-\ui\frac{y_1}{\beta}(k_1+n\uppi/2\Bar{h})}}{\sqrt{\kappa_n-(k_1+n\uppi/2\Bar{h})}}\bigg(E(r_t(\kappa_n-(k_1+n\uppi/2\Bar{h})))-E(r_r(\kappa_n-(k_1+n\uppi/2\Bar{h})))\bigg)  \IEEEeqnarraynumspace\label{eqn:G_s^n}\\
			&&\<-\dfrac{(-1)^n\ue^{-\ui\frac{y_1}{\beta}(k_1-n\uppi/2\Bar{h})}}{\sqrt{\kappa_n-(k_1-n\uppi/2\Bar{h})}}\bigg(E(r_t(\kappa_n-(k_1-n\uppi/2\Bar{h})))-E(r_r(\kappa_n-(k_1-n\uppi/2\Bar{h})))\bigg) \bigg].\IEEEnonumber
	\end{IEEEeqnarray}
	
    When the location is downstream of the serration root, $I(kr_r,\theta_r;\Theta)$ vanishes according to equation~\eqref{eqn:I_Fresnel},  hence $H_n(r_r,\theta_r)=0$. Therefore, the second terms in the square brackets in equation~\eqref{eqn:G_s^n} are equal to 0. Hence, the $n$th scattered surface pressure ($n\neq 0$) downstream of the serration root reduces to
	\begin{IEEEeqnarray}{rl}
			G_s^{(n)}(y)= \dfrac{1-\ui}{2n\uppi}\ue^{\ui\chi_n y_2}\ue^{-\ui kM_0y_1/\beta^2}&\sqrt{k_1-\kappa_n}\bigg[ \dfrac{\ue^{-\ui\frac{y_1}{\beta}(k_1+n\uppi/2\Bar{h})}}{\sqrt{\kappa_n-(k_1+n\uppi/2\Bar{h})}}E(r_t(\kappa_n-(k_1+n\uppi/2\Bar{h})))  \IEEEnonumber\\
			&\<-(-1)^n\dfrac{\ue^{-\ui\frac{y_1}{\beta}(k_1-n\uppi/2\Bar{h})}}{\sqrt{\kappa_n-(k_1-n\uppi/2\Bar{h})}}E(r_t(\kappa_n-(k_1-n\uppi/2\Bar{h}))) \bigg] .\IEEEeqnarraynumspace
		\label{eqn:G_s^n_2}
	\end{IEEEeqnarray}
	
	As mentioned earlier, when $n=0$, function $H_n$ takes a different form of
	\begin{IEEEeqnarray}{rll}
			H_0(r,\theta) &=& \dfrac{\ui\uppi}{4\sqrt{2\kappa_0}\kappa_0 \Bar{h}}\dfrac{1}{\sin^2\frac{\Theta_0}{2}}\bigg[\dfrac{I(\kappa_0 r.\theta;\Theta_0)}{\sin\frac{\Theta_0}{2}}-\dfrac{(2\ui \kappa_0 r)J(\kappa_0 r.\theta;\Theta_0)}{\cos\frac{\Theta_0}{2}}-\dfrac{\ue^{-\ui\frac{\uppi}{4}}}{\sqrt{\uppi}}2\sqrt{2\kappa_0 r}\sin\dfrac{\theta}{2}\ue^{\ui \kappa_0 r}\bigg],\IEEEeqnarraynumspace
		\label{eqn:H_0_pre}
	\end{IEEEeqnarray}
	where the function $J$ is defined by 
	\begin{IEEEeqnarray}{rll}
			J(\kappa_0 r,\theta;\Theta_0) = \dfrac{\ue^{-\ui\frac{\uppi}{4}}}{\sqrt{\uppi}}\bigg[&\sin(\Theta_0+\theta)\ue^{-\ui\kappa_0 r\cos(\Theta_0+\theta)}\Bar{F}\left(\sqrt{2\kappa_0 r}\cos\dfrac{\Theta_0+\theta}{2}\right)\IEEEnonumber\\
			&-\sin(\Theta_0-\theta)\ue^{-\ui\kappa_0 r\cos(\Theta_0-\theta)}\Bar{F}\left(\sqrt{2\kappa_0 r}\cos\dfrac{\Theta_0-\theta}{2}\right)\bigg].\IEEEeqnarraynumspace
		\label{eqn:J}
	\end{IEEEeqnarray}
	When $\theta=0$, $J(\kappa_0 r,0;\Theta_0)$ vanishes.
    To calculate $G_s^{(0)}$, we only need to 
	examine the $\theta=\uppi$, in which case $J$ in equation~\eqref{eqn:J} reduces to
	\begin{equation}
		\begin{aligned}
			J(\kappa_0 r,\uppi;\Theta_0) &= -\sin\Theta_0 (1-\ui)\ue^{\ui \kappa_0 r\cos\Theta_0}E(\kappa_0 r(1-\cos\Theta_0)).
		\end{aligned}
	\end{equation}
	Letting $\cos\Theta_0=k_1/\kappa_0$ 
    and $\theta=\uppi$, we can re-organise equation~\eqref{eqn:H_0_pre} to be
	\begin{IEEEeqnarray*}{rl}
			H_0(r,\uppi) &=\dfrac{\ui\uppi}{2\sqrt{\kappa_0} \Bar{h}}\dfrac{1-\ui}{\kappa_0-k_1}\times\\
			&\bigg[\bigg(\dfrac{1}{\sqrt{1-k_1/\kappa_0}}+2\ui\kappa_0 r\sqrt{1-k_1/\kappa_0}\bigg)\ue^{\ui k_1r}E(\kappa_0 r(1-k_1/\kappa_0))-\sqrt{\dfrac{2}{\uppi}}\sqrt{\kappa_0 r}\ue^{\ui \kappa_0 r}\bigg].\IEEEeqnarraynumspace\IEEEyesnumber
		\label{eqn:H_0}
	\end{IEEEeqnarray*}
	
	When the location is upstream of the serration root, substituting equation~\eqref{eqn:H_0} into \eqref{eqn:G_s} we obtain
	\begin{IEEEeqnarray*}{rll}
			G_s^{(0)}(y)&=\dfrac{\ue^{-\ui kM_0y_1/\beta^2}\ue^{\ui\chi_0y_2}}{4\sqrt{\kappa_0(k_1-\kappa_0)}\Bar{h}}\bigg\{ &\dfrac{1-\ui}{\sqrt{1-k_1/\kappa_0}}\ue^{\ui k_1r}\bigg[E(r_t(\kappa_0-k_1))-E(r_r(\kappa_0-k_1))\bigg] \\
			&&\<+2(\ui+1)\kappa_0\sqrt{1-k_1/\kappa_0}\ue^{\ui k_1r}r\bigg[E(r_t(\kappa_0-k_1))-E(r_r(\kappa_0-k_1))\bigg]\\
			&&\<+2(\ui+1)\kappa_0\sqrt{1-k_1/\kappa_0}\ue^{\ui k_1 r}\Bar{h}\bigg[E(r_t(\kappa_0-k_1))+E(r_r(\kappa_0-k_1))\bigg]\\
			&&\<-\dfrac{\ue^{-\ui\frac{\uppi}{4}}}{\sqrt{\uppi}}2\sqrt{\kappa_0 r_t}\ue^{\ui \kappa_0 r_t}\ue^{-\ui k_1\Bar{h}}+\dfrac{\ue^{-\ui\frac{\uppi}{4}}}{\sqrt{\uppi}}2\sqrt{\kappa_0 r_r}\ue^{\ui \kappa_0 r_r}\ue^{\ui k_1\Bar{h}}\bigg\}.\IEEEeqnarraynumspace\IEEEyesnumber\label{eqn:G_s_0_1}
	\end{IEEEeqnarray*}
	Similar to equation~\eqref{eqn:G_s^n_2}, when the location is downstream of the serration root, $G_s^{(0)}(y)$ takes the form of
	\begin{IEEEeqnarray*}{rll}
			G_s^{(0)}(y)
			&=\dfrac{\ue^{-\ui kM_0y_1/\beta^2}\ue^{\ui\chi_0y_2}}{4\sqrt{\kappa_0(k_1-\kappa_0)}\Bar{h}}\bigg\{ &\dfrac{1-\ui}{\sqrt{1-k_1/\kappa_0}}\ue^{\ui k_1r}E(r_t(\kappa_0-k_1))-\dfrac{\ue^{-\ui\frac{\uppi}{4}}}{\sqrt{\uppi}}2\sqrt{\kappa_0 r_t}\ue^{\ui \kappa_0 r_t}\ue^{-\ui k_1\Bar{h}}\IEEEeqnarraynumspace \\
			&&\<+2(\ui+1)\kappa_0\sqrt{1-k_1/\kappa_0}\ue^{\ui k_1r}r E(r_t(\kappa_0-k_1))\IEEEeqnarraynumspace\IEEEyesnumber\label{eqn:G_s_0}\\
			&&\<+2(\ui+1)\kappa_0\sqrt{1-k_1/\kappa_0}\ue^{\ui k_1 r}\Bar{h}E(r_t(\kappa_0-k_1))\bigg\}.
	\end{IEEEeqnarray*}

    Thus, we obtain the full explicit expression for the scattered pressure 
    \begin{equation}
	 	G_s=\sum_{n=-\infty}^{\infty}G_s^{(n)}.
	  \end{equation}
    Next, we will integrate this scattered pressure over the plate surface to obtain the far-field sound pressure $p_f$. Since we assumed $P_i=1$, $p_f$ represents the transfer function between the input pressure fluctuation and the far-field sound. 
	
	\subsection{The far-field sound pressure spectrum}
	\label{section: The far-field sound pressure spectrum}
	The far-field sound can be obtained using the surface pressure integral, as described in Amiet's model\citep{amiet_noise_1976}. When the observation point is at $(x_1,x_2,x_3)$, we can obtain the far-field sound pressure $p_f$ through the Curle's integral, i.e.
	\begin{equation}
		p_f(\boldsymbol{x},\omega) = \dfrac{-\ui\omega x_3}{4\uppi c_0S_0^2}\int_{-N-\frac{1}{4}}^{N+\frac{3}{4}}\int_{-c}^{hF(y_2)}\Delta P(y_1,y_2)\ue^{-\ui kR}\ud y_1 \ud y_2,
		\label{eqn:pf_pre1}
	\end{equation}
	where the outer integral from $-N - 1/4$ to $N + 3/4$ represents $2N + 1$ serrations along the trailing edge of the flat plate, and $\Delta P = G_s + P_{in}$ denotes the pressure jump across the surface, $S_0^2 = x_1^2 + \beta^2(x_2^2+x_3^2)$ and 
	\begin{equation}
		R=\dfrac{M(x_1-y_1)-S_0}{\beta^2}+\dfrac{x_1y_1+x_2y_2\beta^2}{\beta^2S_0}.
	\end{equation}
	Substituting the equations \eqref{eqn:G_s^n}, \eqref{eqn:G_s^n_2}, \eqref{eqn:G_s_0_1} and \eqref{eqn:G_s_0} into equation~\eqref{eqn:pf_pre1} yields the far-field sound pressure $p_f$.
	When $n\neq 0$, the $n$th component of $p_f$, i.e. $p_f^{(n)}$, can be formulated as 
	\begin{IEEEeqnarray*}{rll}
			p_f^{(n)} &= \dfrac{\ui\omega x_3}{4\uppi c_0}\sum_{m=-N}^{m=N}\sum_{j}\bigg[&D_{nj}\int_{m-\frac{1}{4}}^{m+\frac{3}{4}}\ue^{-\ui \Omega_ny_2}\int_{-c}^{hF(y_2)}\ue^{-\ui \sigma_{nj}y_1}E(\gamma_{nj}(-y_1+h))\ud y_1\ud y_2\\
			&&\<+D_{nj}\int_{m-\frac{1}{4}}^{m+\frac{3}{4}}\ue^{-\ui \Omega_ny_2}\int_{-c}^{-h}\ue^{-\ui \sigma_{nj}y_1}E(\gamma_{nj}(-y_1-h))\ud y_1\ud y_2\bigg],\IEEEyesnumber
		\label{eqn:pf_n}
	\end{IEEEeqnarray*}
	where the coefficients are defined as
	\begin{IEEEeqnarray*}{rCl}
			\sigma_{n1} &=& \dfrac{k_1+n\uppi/2\uph}{\beta}+\dfrac{kx_1}{\beta^2S_0},\\
			\sigma_{n2} &=& \dfrac{k_1-n\uppi/2\uph}{\beta}+\dfrac{kx_1}{\beta^2S_0},\\
			\gamma_{n1} &=& \dfrac{\kappa_n-(k_1+n\uppi/2\uph)}{\beta},\\
			\gamma_{n2} &=& \dfrac{\kappa_n-(k_1-n\uppi/2\uph)}{\beta},\IEEEyesnumber\\
			D_{n1} &=& \dfrac{-1}{S_0^2}\uexp\bigg\{-\ui k\dfrac{M_0x_1-S_0}{\beta^2}\bigg\}\dfrac{\sqrt{k_1-\kappa_n}}{\kappa_n-(k_1+n\uppi/2\uph)}\dfrac{1-\ui}{2n\uppi},\\
			D_{n2} &=& (-1)^n\dfrac{1}{S_0^2}\uexp\bigg\{-\ui k\dfrac{M_0x_1-S_0}{\beta^2}\bigg\}\dfrac{\sqrt{k_1-\kappa_n}}{\kappa_n-(k_1-n\uppi/2\uph)}\dfrac{1-\ui}{2n\uppi},\\
			\Omega_n &=& -\chi_n+k\dfrac{x_2}{S_0}.
	\end{IEEEeqnarray*}

    Equation~\eqref{eqn:pf_n} can be evaluated by using integration by parts to yield
    \begin{IEEEeqnarray}{l}
        p_f^{(n)} = \dfrac{\sin(\Omega_0(N+\frac{1}{2}))}{\sin{(\Omega_0/2)}}\dfrac{\ui\omega x_3}{4\uppi c_0}g^{(n)}.
    \end{IEEEeqnarray}
    The non-dimensionalised scattered noise $g^{(n)}$ is evaluated as
    \begin{IEEEeqnarray*}{rll}
			g^{(n)}&=\sum_{j=1,2}&\dfrac{\ui D_{nj}\ue^{-\ui \Omega_n/4}}{\sigma_{nj}}\dfrac{2}{\Omega_n}\sin{\dfrac{\Omega_n}{2}}
            \bigl\{\pm\ue^{\mp\ui \sigma_{nj}h}\Q\left(2\gamma_{nj},2\sigma_{nj},c\pm h\right)\mp\ue^{\ui \sigma_{nj}c}\Q\left(2\gamma_{nj},0,c\pm h\right)\bigr\}\\
			&&\<+\dfrac{4D_{nj}h\ue^{-\ui \Omega_n/4}\ue^{-\ui \sigma_{nj}h}}{\Omega_n(\Omega_n\pm4\sigma_{nj}h)}\Q\left(4\gamma_{nj},4\sigma_{nj}\pm \Omega_n/h,h\right)\IEEEeqnarraynumspace\IEEEyesnumber\label{eqn:nneq0_curle}\\
            &&\<\pm\dfrac{D_{nj}\ue^{-\ui \Omega_n/4}}{\sigma_{nj}}\left\{\dfrac{\ue^{\pm\ui \frac{\Omega_n}{2}}\ue^{\ui \sigma_{nj}h}}{\Omega_n\pm4\sigma_{nj}h}\Q\left(4\gamma_{nj},0,h\right)+\dfrac{\ue^{\mp\ui \frac{\Omega_n}{2}}\ue^{-\ui \sigma_{nj}h}}{\Omega_n}\Q(4\gamma_{nj},4\sigma_{nj},h)\right\},\IEEEeqnarraynumspace
	\end{IEEEeqnarray*}
    where
    \begin{IEEEeqnarray}{l}
        \Q(p,q,x) = \sqrt{\dfrac{p}{p+q}}E\left(\dfrac{1}{2}(p+q)x\right).
    \end{IEEEeqnarray}
    The symbols $\pm$ and $\mp$ in the same line mean that both the positive and negative terms appear in the summand. The same notation will be used in the integral results for $n=0$.

	In a similar manner, the $0$th component of the far-field sound $p_f$ can be obtained and written as a sum over $j$, i.e. $p_f^{(0)} = \sum_j p_f^{(0,j)}$. It should be noted that $j$ takes the value from 1 to 4 when $n=0$, while it takes the value from 1 to 2 for $n\neq 0$. The detailed formula for $p^{(0,j)}_f$ can be found in Appendix~\ref{appendix:n0}.
	
	Considering the incident gust in equation~\eqref{eqn:P_i}, the far-field sound pressure can be written as
	\begin{IEEEeqnarray*}{rCl}
                p_f&=&\dfrac{\sin(\Omega_0(N+\frac{1}{2}))}{\sin{(\Omega_0/2)}}\left(\dfrac{\ui\omega x_3}{4\uppi c_0}\right)\mathcal{L}(\omega,K_1,K_2)P_i(\omega,K_1,K_2).\IEEEeqnarraynumspace\IEEEyesnumber
		\label{result:p_f}
	\end{IEEEeqnarray*}
	$\mathcal{L}(\omega,K_1,K_2)$ is the non-dimensionalised transfer function from the incident gust to the far-field sound pressure, which is defined as 
    \begin{IEEEeqnarray*}{rCl}
                \mathcal{L}(\omega,K_1,K_2)&=&\sum_{n\neq 0}g^{(n)} + \sum_{j}g^{(0,j)},\IEEEeqnarraynumspace\IEEEyesnumber
                \label{result:L}
	\end{IEEEeqnarray*}
	
	A recent study of the authors~\citep{Tian_Lyu_2024} shows that at high frequencies the non-frozen nature of the wall pressure fluctuations beneath turbulent boundary layers has a significant impact on the noise emissions from serrated trailing edges. However, such effects may be accounted for by modifying the convection velocity in models based on frozen turbulence. Therefore, this paper still assumes frozen turbulence. Doing so also enables a direct comparison with earlier two-dimensional models. Under the frozen assumption, the surface pressure fluctuation beneath the turbulent boundary layer peaks in the vicinity of the convective wavenumber $\omega/U_c$. Thus, the hypothetical surface pressure fluctuation of frequency $\omega$ can be expressed as a Fourier integral, i.e.
	\begin{IEEEeqnarray}{rCl}
		P_{in}(\omega,y_1,y_2)&=&\int_{-\infty}^{\infty}\int_{-\infty}^{\infty}P_i(\omega,K_1,K_2)\ue^{\ui (K_1y_1+K_2y_2)}\ud K_1\ud K_2,
	\end{IEEEeqnarray}
	where the integral for $K_1$ is simplified as an integral over a decomposed Dirac delta function, i.e.
    \begin{IEEEeqnarray}{rCl}
		P_{in}(\omega,y_1,y_2)&=&\int_{-\infty}^{\infty}P_i(\omega,K_2)\ue^{\ui (\omega y_1/U_c+K_2y_2)}\ud K_2.
		\label{eqn:P_incident}
	\end{IEEEeqnarray}
    Thus, the wall pressure defined in equation~\eqref{eqn:P_incident} induces a far-field sound pressure $p_f(\boldsymbol{{x}, \omega)}$ at the position $\boldsymbol{x}$ and frequency $\omega$, i.e.
	\begin{IEEEeqnarray}{rCl}
		p_f(\boldsymbol{x}, \omega) =\left(\dfrac{\ui\omega x_3}{4\uppi c_0}\right)\int_{-\infty}^{\infty}\left(\dfrac{\sin(\Omega_0(N+\frac{1}{2}))}{\sin{\frac{\Omega_0}{2}}}\right)\mathcal{L}\left(\omega,\frac{\omega}{U_c},K_2\right)P_i(\omega,K_2)\ud K_2.
	\end{IEEEeqnarray}
	
	The power spectrum density (PSD) of $p_f$ can be calculated as
	\begin{IEEEeqnarray}{rCl}
		S_{pp}=\lim_{T\rightarrow\infty}\dfrac{\uppi}{T}\langle p_f(\boldsymbol{x},\omega)p_f^*(\boldsymbol{x},\omega)\rangle,
        \label{eqn:PSD}
	\end{IEEEeqnarray}
where the bracket $\langle \rangle$ represents the ensemble average. When the span is considered large compared to the serration wavelength, $N$ obtains a large number. We can use the approximation that as $N\to\infty$
	\begin{IEEEeqnarray}{l}
		\lim_{N\rightarrow\infty}\left(\dfrac{\sin(\Omega_0(N+\frac{1}{2}))}{\sin{\frac{\Omega_0}{2}}}\right)^2\sim 2\uppi\D\sum_{m=-\infty}^{\infty}\delta\left(K_2+\frac{kx_2}{S_0}+2m\uppi\right),\IEEEeqnarraynumspace
	\end{IEEEeqnarray}
	where $\D$ denotes the span of the plate, while $m$ denotes the spanwise mode number. When the wavenumber spectral density of the incident wall pressure fluctuations is given as $\Pi\left(\omega,kx_2/S_0 +2m\uppi\right)$, $S_{pp}$ can be simplified as
	\begin{IEEEeqnarray}{rCl}
		S_{pp}=\left(\dfrac{\omega x_3}{4\uppi c_0}\right)^22\uppi\D\sum_{m=-\infty}^{\infty}\left|\mathcal{L}\left(\omega,\frac{\omega}{U_c},\frac{kx_2}{S_0}+2m\uppi\right)\right|^2 \Pi\left(\omega,\frac{kx_2}{S_0}+2m\uppi\right).
		\label{result_Spp}
	\end{IEEEeqnarray}
    Equations~\eqref{result:L} and~\eqref{result_Spp} serve as the principal formulations for predicting the far-field sound spectra.
    	
    \section{Result and Discussion}\label{sec:Model Validation}
    Equation~\eqref{result_Spp} relates the far-field sound pressure spectrum to the wall pressure wavenumber-frequency spectrum. Once this wavenumber-frequency spectrum is known, the power spectral density of the far-field sound can be readily calculated using (\ref{result_Spp}). In this section, we first show the scattered pressure calculated in \ref{sec:scatter} as a validation. Far-field noise predictions are then computed and compared with the two-dimensional Wiener-Hopf method-based model. Finally, the computational efficiency of the present model is assessed by benchmarking its performance against Lyu’s iterative Schwarzschild-based three-dimensional model.

	\subsection{Distribution of the scattered pressure}\label{sec:scatter}
	We begin by examining the form of $G_s$ in the limit $h\rightarrow0$. As $h$ approaches 0, the only nonzero term is the third term in equation~\eqref{eqn:G_s_0_1}, i.e.
	\begin{IEEEeqnarray}{rll}
			G_s^{(0)}(y)\bigg|_{h\rightarrow 0}&=&\dfrac{\ue^{-\ui kM_0y_1/\beta^2}\ue^{\ui\chi_0y_2}}{2\sqrt{\kappa_0(k_1-\kappa_0)}}(\ui+1)\kappa_0\sqrt{1-k_1/\kappa_0}\ue^{\ui k_1 r}\bigg[E(r_t(\kappa_0-k_1))+E(r_r(\kappa_0-k_1))\bigg].\IEEEeqnarraynumspace
	\end{IEEEeqnarray}
	When the geometric reflection wave $P_r$ is removed, this equation precisely reduces to the straight-edge solution presented in \citet{amiet_noise_1976}. 

    The Green's function in \citet{lyu_analytical_2023} is obtained by decomposing the acoustic pressure into three components: the incident wave $P_{in}$, the geometrically reflected wave $P_r$, and the reflection-removed scattered field $R_s$. The reflected wave $P_r$ is straightforward to obtain and takes the form of
	\begin{equation}
		P_{r}=P_i(\omega,K_1,K_2)\ue^{-\ui\omega t}\ue^{-\ui\left(K_1y_1+K_2y_2-K_3y_3\right)},
	\end{equation}
    and
    \begin{IEEEeqnarray}{rCl}
        R_s = G_s - P_r.
        \label{eqn:R_s}
    \end{IEEEeqnarray}

\begin{figure}
		\centerline{\includegraphics[width=0.7\textwidth]{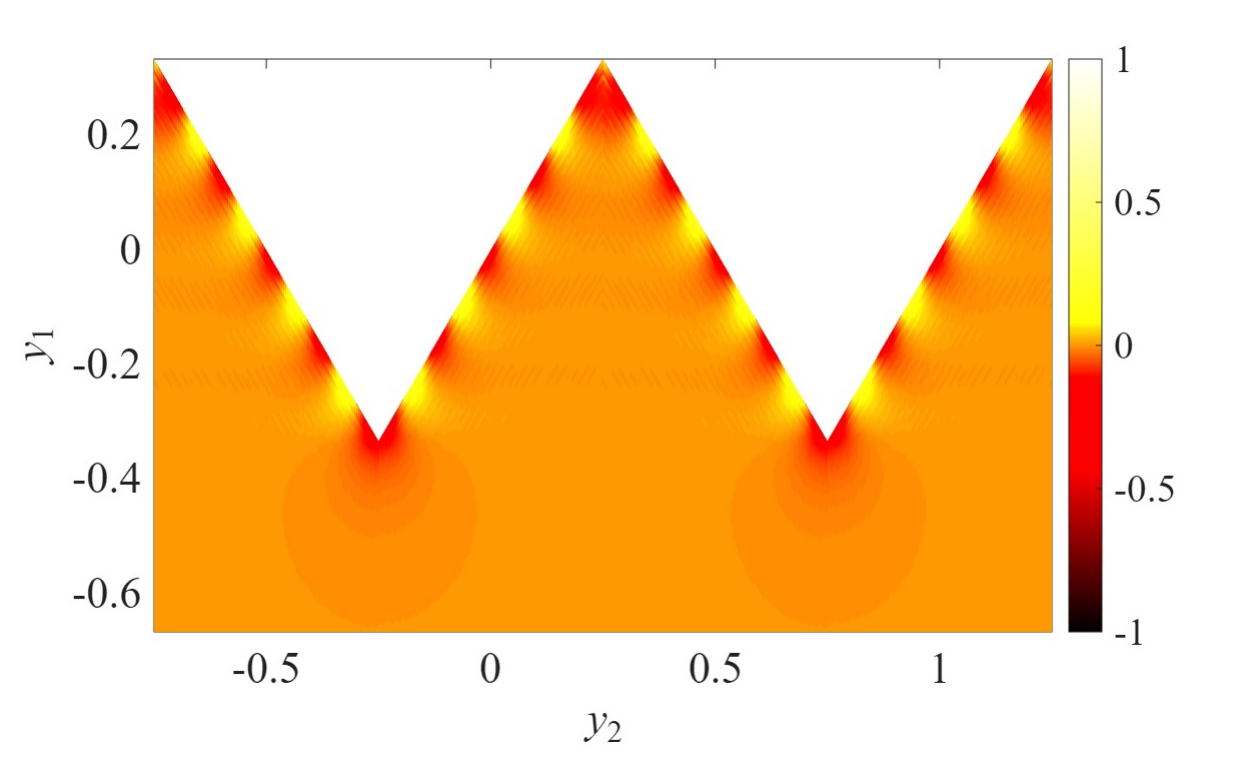}}
		\caption{Distribution of the real part of the scattered pressure $R_s$ in equation~\eqref{eqn:R_s}. $R_s$ is resulted from a wall pressure gust with $K_2=0$, $M_0=0.1$ and $h=1/3$.}
		\label{fig:surf_p}
	\end{figure}

	When $h$ obtains a finite value, the distribution of $R_s$ in equation~\eqref{eqn:R_s} is shown in figure~\ref{fig:surf_p}. This distribution closely resembles that presented in \citet{lyu_prediction_2016}. The reflection-removed scattered pressure is concentrated in a narrow strip along the edge. As the frequency increases, the strip becomes progressively narrower. When the frequency is sufficiently high, the wavelength of the incident wave is much smaller than the characteristic length of the serrated edge. Each serration locally tends to behave like a separate straight line, resulting in the scattered field similar to that shown in figure \ref{fig:surf_p}. 
    
    \subsection{Comparison with the two-dimensional model based on the Wiener-Hopf method}
    To show that the present model successfully extends to three-dimensional, we perform a systematic comparison with the earlier two-dimensional model based on the Wiener-Hopf technique. The principal distinction between these models lies in the predicted decay of sound pressure amplitude with the observer $r$, namely $1/r$ for the three-dimensional model versus $1/\sqrt{r}$ for the two-dimensional model. For comparison, we set $x_3/c=1$. At this relatively close position to the aerofoil, we may expect the sound field to be approximately two-dimensional; as such, the three-dimensional prediction should align fairly closely with the two-dimensional predictions.  

    We compare the predicted far-field sound power spectral density, $S_{pp}$ shown in \eqref{result_Spp}, as a function of frequency. For illustrative purposes, we use Chase’s wavenumber-frequency spectral model \citep{chase1987character} similar to previous studies~\citep{lyu_prediction_2016}. According to Chase, the wavenumber-frequency spectrum can be modelled as
	\begin{IEEEeqnarray}{rCl}
        \Pi(\omega, k_1, k_2)
        = \frac{C_m\,\rho_0^2\,\nu_*^3\,k_1^2\,\delta^5}
            {\bigl[(k_1 - \omega/U_c)^2 (\delta\,U_c\,\nu_*/3)^2 + (k_1^2 + k_2^2)\,\delta^2 + \chi^2\bigr]^{5/2}},
        \label{eqn:wall_Pi}
    \end{IEEEeqnarray}
    where $U_c$ can be approximated by $U_c = 0.7\,U$, $\rho_0$ is the fluid density and the empirical coefficients are $C_m \approx 0.1553$, $\chi \approx 1.33$, and $\nu_* \approx 0.03\,U$. The turbulent boundary layer thickness, $\delta$, is estimated by
	\begin{IEEEeqnarray}{rCl}
        \frac{\delta}{c} = 0.382\,Re_c^{-1/5},
        \label{eqn:tbl_delta}
    \end{IEEEeqnarray}
    with $Re_c$ denoting the chord-based Reynolds number.  
    Since the wavenumber spectrum exhibits a dominant peak near $k_1 = \omega/U_c$, one often integrates \eqref{eqn:wall_Pi} over $k_1$, retaining only the leading-order contributions \citep{howe_noise_1991} to obtain
	\begin{IEEEeqnarray}{rCl}
        \Pi(\omega, k_2)
        \approx \frac{4\,C_m\,\rho_0^2\,\nu_*^4\,(\omega/U_c)^2\,\delta^4}
           {U_c \bigl[(\omega/U_c)^2 + k_2^2)\,\delta^2 + \chi^2\bigr]^2}.
        \label{eqn:wall_Pi_simplified}
    \end{IEEEeqnarray}

    \begin{figure}
		\centerline{\includegraphics[width=0.8\textwidth]{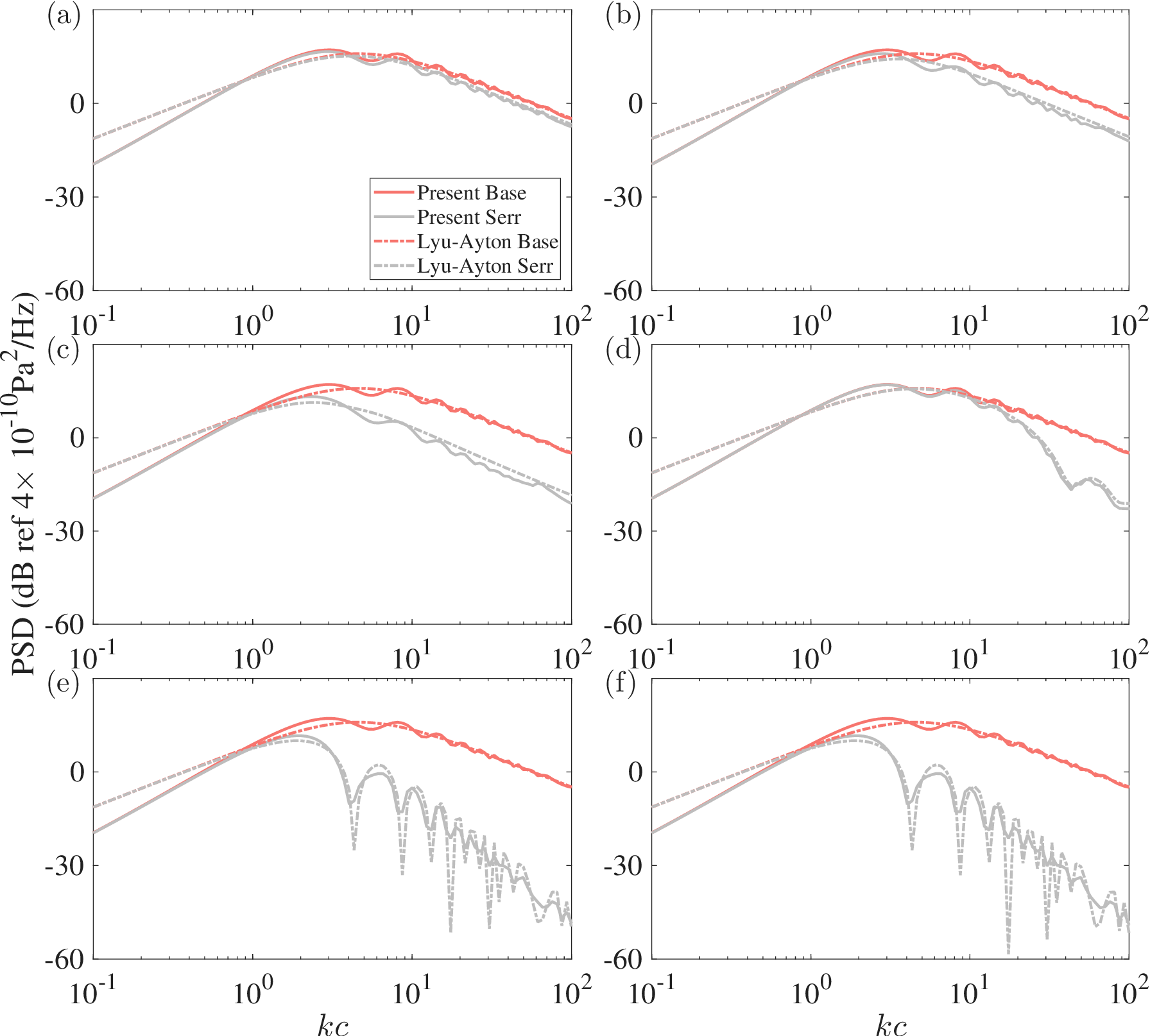}}
		\caption{Comparison of the PSD of the predicted noise from the two- and three-dimensional models for serrated and straight trailing edges. The observer at $\theta = 90^\circ$ and $x_3/c=1$ above the plate in the mid-span plane with $M_0=0.1$: (a) $\lambda/h=8, h/c=0.025$; (b) $\lambda/h=4, h/c=0.025$; (c) $\lambda/h=2, h/c=0.05$; (d) $\lambda/h=1, h/c=0.005$; (e) $\lambda/h=0.4, h/c=0.05$; (f) $\lambda/h=0.2, h/c=0.05$.}
		\label{fig:PSD_compare}
    \end{figure}
    Using this integrated Chase model as input, we are now in a position to compare the PSD \eqref{eqn:PSD} of the predicted trailing‐edge noise from the present and earlier two-dimensional models under varying serration profiles. 
    As shown in Figure~\ref{fig:PSD_compare}, the PSD predicted by both models exhibits similar trends across all parameters. At low frequencies, the spectra corresponding to the straight and serrated trailing edges nearly collapse in both models, indicating that the base and straight edges generate comparable noise levels. However, the two-dimensional model consistently predicts higher noise than the present three-dimensional model in this regime. This discrepancy arises because the two-dimensional model assumes a semi-infinite plate, which produces a higher low-frequency component. In contrast, the three-dimensional model accounts for the finite span and streamwise extent of the plate, resulting in dipolar directivity and a correspondingly lower sound amplitude at low frequencies.
    In the high-frequency regime, both models predict a pronounced reduction in noise when serrations are used, as shown in Figure~\ref{fig:PSD_compare}. The agreement between the two models is good for both the baseline and serrated configurations. At these higher frequencies, the trailing-edge noise is dominated by the scattered pressure near the serrated edges~\citep{lyu_prediction_2016}. Because the acoustic wavelength becomes small compared with the plate dimensions, the finite flat plate used in the present model approximates a semi-infinite flat plate increasingly well, leading to good agreement with the two-dimensional prediction.

    \begin{figure}
	\centerline{\includegraphics[width=0.8\textwidth]{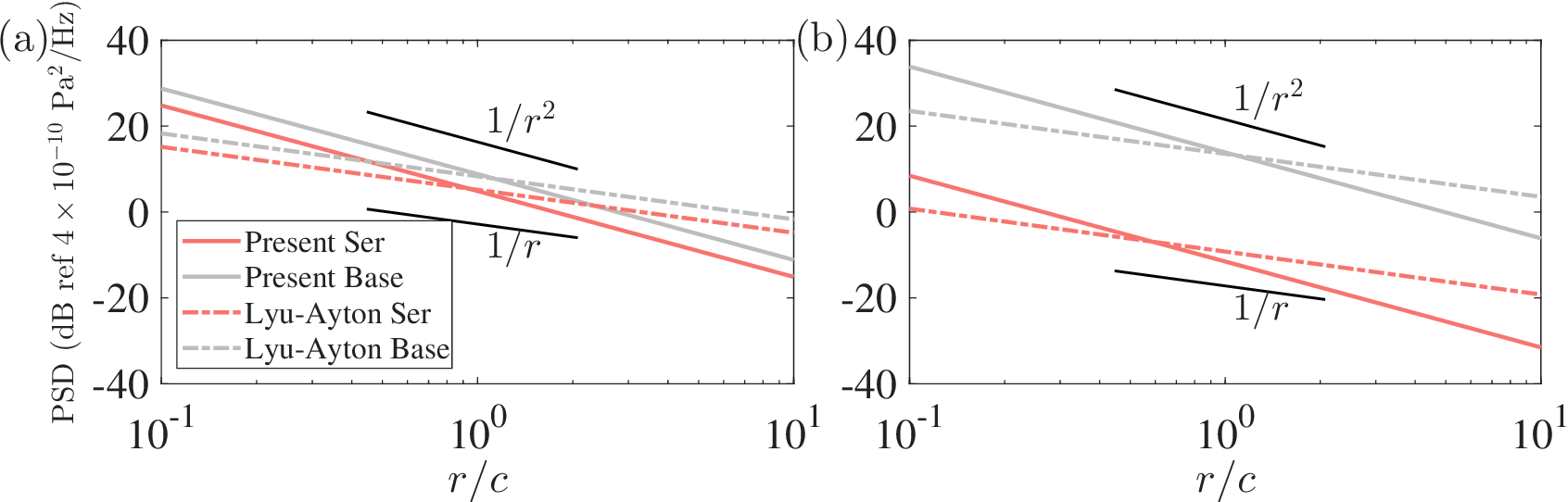}}
	\caption{Comparison of the PSD of the predicted noise from the two- and three-dimensional models for serrated and straight trailing edges. The observer at $\theta = 90^\circ$ and $x_3 = r$ above the plate in the mid-span plane with $M_0=0.1$, $\lambda/h=0.4$, and $h/c=0.1$. (a) $kc=1$; (b) $kc=10$.}
	\label{fig:PSD_compare_2}
    \end{figure}

    The most pronounced manifestation of three-dimensional effects lies in how the predicted scattered sound scales with the observation distance $r$. Figure~\ref{fig:PSD_compare_2} compares the variation of the predicted PSD with~$r$ at several frequencies. The observer is located at $\theta = 90^\circ$ with $x_3 = r$ above the plate in the mid-span plane, and the free-stream Mach number is $M_0 = 0.1$. As expected, the present three-dimensional model predicts a PSD decay proportional to $1/r^2$ across all frequencies, whereas the two-dimensional model yields a slower decay proportional to $1/r$. This contrast reflects the fundamental scaling characteristics of the radiated sound field and confirms that the present model correctly captures the spatial decay behaviour of a three-dimensional acoustic source.

    \begin{figure}
		\centerline{\includegraphics[width=0.8\textwidth]{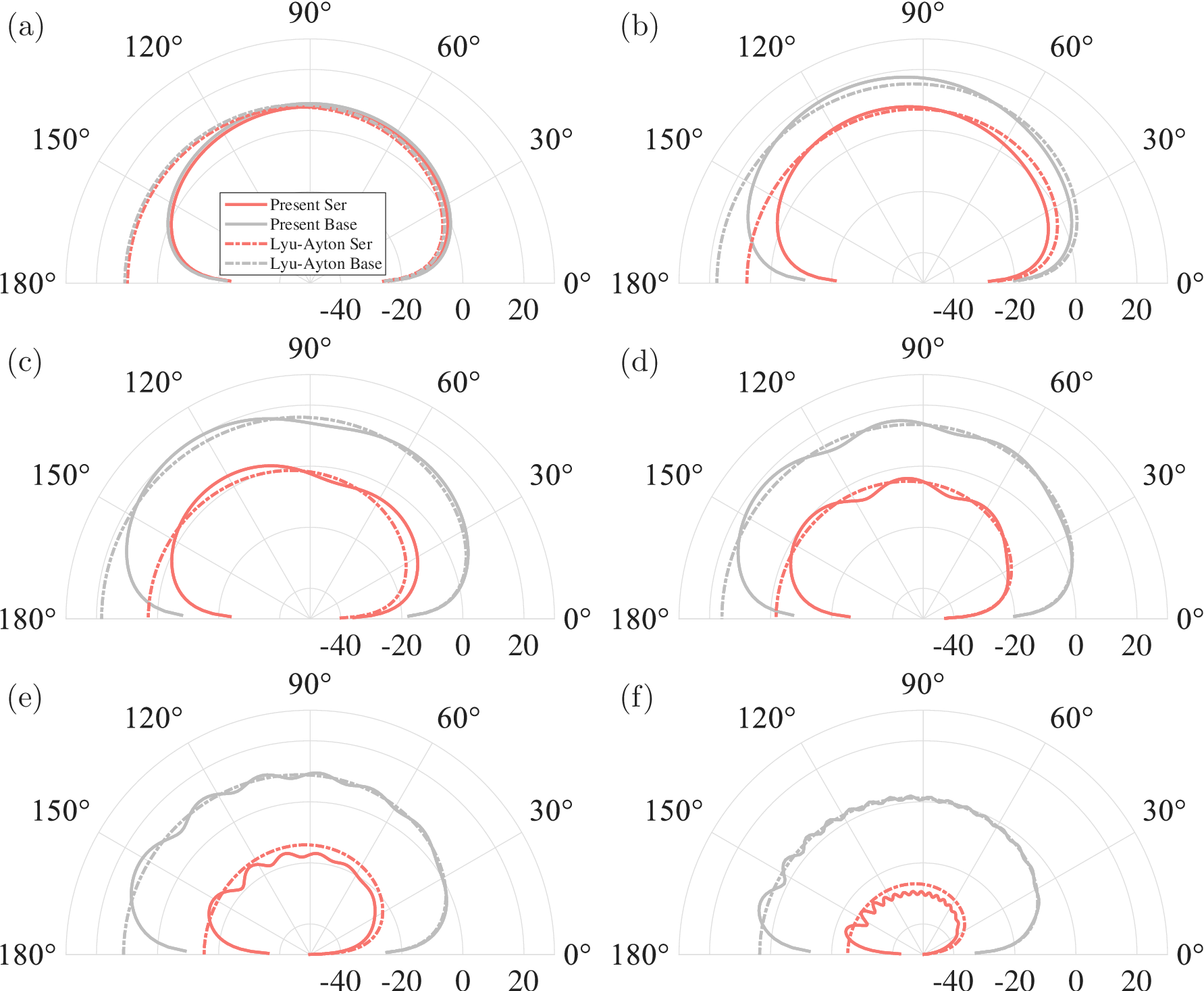}}
		\caption{Comparison of the directivity of the predicted trailing edge noise from the two- and three-dimensional models under different frequencies. The distance between the observer and the trailing edge $r/c=1$ above the plate in the mid-span plane with $\lambda/h=0.4,h/c=0.05$, and $M_0=0.1$. (a) $kc=1$; (b) $kc=3$; (c) $kc=5$; (d) $kc=10$; (e) $kc=20$; (f) $kc=50$.}
		\label{fig:dir_compare1}
    \end{figure}

    In addition to the scaling behaviour, we can also compare the predicted directivity patterns at various frequencies. We again choose the observer to be at $r/c=1$, but now on a circular arc in the mid-span plane. Owing to symmetry, only the directivity pattern in the upper half plane is presented, as can be seen in figure~\ref{fig:dir_compare1}. From figure~\ref{fig:dir_compare1}, one can see close agreement between the two models across virtually the entire observer angle range for both straight and serrated trailing edges. The only difference occurs at large observer angles ($\theta \sim 180^\circ$). This is particularly pronounced at low frequencies, as demonstrated in figure~\ref{fig:dir_compare1}(a–d) for both straight and serrated trailing edges. The three-dimensional model predicts rapid noise attenuation in the upstream direction, a feature absent from the two-dimensional formulation. This discrepancy arises from the finite flat plate used in the present model, resulting in the classical dipolar behaviour at low frequencies, as opposed to cardioid directivity for semi-infinite plates. As frequency increases, this difference becomes increasingly less pronounced, as shown in figures \ref{fig:dir_compare1}(e–f).  In addition, as frequency increases, the present model predicted increasingly pronounced oscillations in the directivity pattern compared to the two-dimensional semi-infinite model. This is expected to be due to the stronger interference effects of the scattered pressure on a finite-size flat plate.

    \begin{figure}
		\centerline{\includegraphics[width=0.8\textwidth]{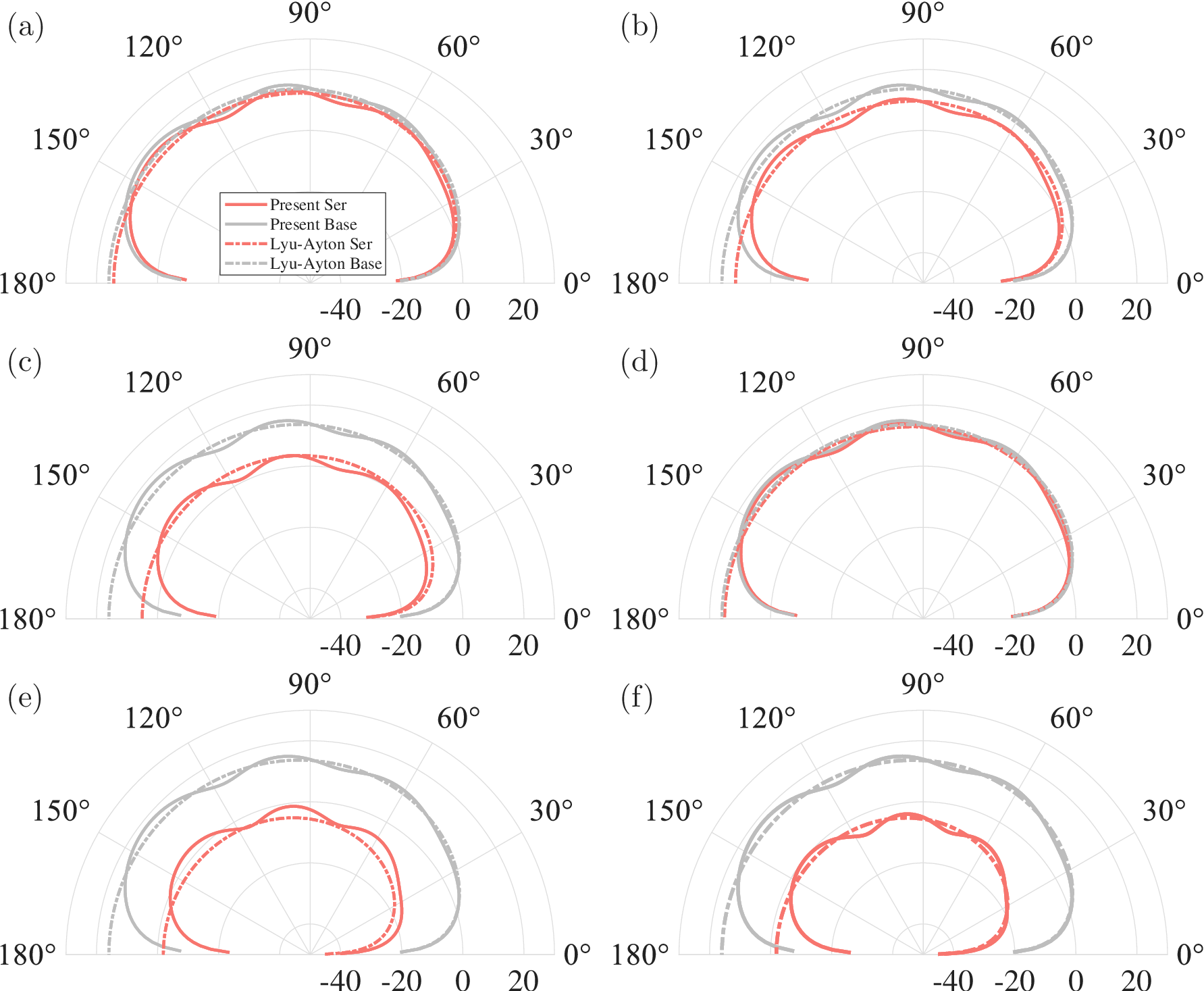}}
		\caption{Comparison of the directivity of the predicted trailing edge noise from the two- and three-dimensional models under different frequencies. The distance between the observer and the trailing edge $r/c=1$ above the plate in the mid-span plane with $M_0=0.1$ and $kc=10$: (a) $\lambda/h=8,h/c=0.025$; (b) $\lambda/h=4,h/c=0.025$; (c) $\lambda/h=2,h/c=0.05$; (d) $\lambda/h=1,h/c=0.005$; (e) $\lambda/h=0.4,h/c=0.05$; (f) $\lambda/h=0.2,h/c=0.05$.}
		\label{fig:dir_compare2}
    \end{figure}
    
    To further examine the robustness of the present model, figure~\ref{fig:dir_compare2} compares the directivity patterns predicted by the two models for various trailing‐edge serration geometries at a dimensionless frequency of $kc=10$ and a convective Mach number of $M_0=0.1$. Observer locations are the same those as in Figure~\ref{fig:dir_compare1}. The comparison shows good agreement between the three‐dimensional and two‐dimensional predictions across all serration configurations. Again, the present model reveals additional directivity lobes due to the interference of the scattered pressure on the finite flat plate. In addition, these lobes appear increasingly pronounced for sharp serrated edges, as shown by figures~\ref{fig:dir_compare2}(e) and (f).

    Figures~\ref{fig:PSD_compare}-\ref{fig:PSD_compare_2} demonstrate that the present model not only recovers the spectra and directivity patterns predicted by the two-dimensional model but also captures key three-dimensional characteristics. These include the correct $1/r^{2}$ decay of the sound pressure PSD, the vanishing radiation at $\theta=180^\circ$, and the directivity oscillations induced by the finite span and chord of the flat plate. Accurately capturing these features is essential for practical applications, particularly those involving rotating blade elements, where the observer–blade distance and orientation vary continuously during operation.

    \subsection{Computational cost}
	In this section, we discuss the computational efficiency of the present model and its comparison with earlier Schwarzschild-based models. The present model is written as a double infinite summation, therefore computational efficiency is expected to improve considerably. We benchmark both models in MATLAB in predicting noise spectra using 61 spanwise modes ($m=30$). In each spectrum 100 frequencies are chosen. The following four sets of parameters are used: (a)$M=0.1,$ $\lambda/h=6,$ $h/c=0.025;$ (b)$M=0.2,$ $\lambda/h=6,$ $h/c=0.025;$ (c)$M=0.1,$ $\lambda/h=3,$ $h/c=0.05;$ (d)$M=0.1,$ $\lambda/h=2,$ $h/c=0.05$. A Standard PC equipped with an intel i9-13900KF and 64GB of RAM is used.
	\begin{table}[!htbp]
		\begin{center}
			\def~{\hphantom{0}}
			\begin{tabular}{ccccccc}
                    \hline
				~~~	& \multicolumn{2}{c}{The present model} & \multicolumn{2}{c}{Lyu's model} & \multicolumn{2}{c}{
					speed-up ratio}\\[3pt]
				(a)   & \multicolumn{2}{c}{1.2656~} & \multicolumn{2}{c}{41.8906} & \multicolumn{2}{c}{33.0994} \\
				(b)   & \multicolumn{2}{c}{0.6875~} & \multicolumn{2}{c}{36.7188} & \multicolumn{2}{c}{53.4092} \\
				(c)   & \multicolumn{2}{c}{0.95312} & \multicolumn{2}{c}{33.9219} & \multicolumn{2}{c}{35.5904} \\
				(d)   & \multicolumn{2}{c}{0.9375~} & \multicolumn{2}{c}{45.4531} & \multicolumn{2}{c}{48.4833} \\
                    \hline
			\end{tabular}
			\caption{Computation time (s) of the two models for 100 frequency points and 61 spanwise modes ($m=30$). The flow and geometric parameters: (a) $M=0.1,$ $\lambda/h=6,$ $h/c=0.025;$ (b) $M=0.2,$ $\lambda/h=6,$ $h/c=0.025;$ (c) $M=0.1,$ $\lambda/h=3,$ $h/c=0.05;$ (d) $M=0.1,$ $\lambda/h=2,$ $h/c=0.05.$}
			\label{tab:time cost}
		\end{center}
	\end{table}
	
	Table \ref{tab:time cost} presents the computing cost of the two models. From the table, one can see that the present model is significantly faster than the previous Schwarzschild-based iterative model. In general, the elapsed time of the present model is less than 3\% of that used by the iterative model. This two orders of magnitude efficiency improvement holds a significant potential for serration shape optimisations. 
    It is important to emphasise that, as mentioned in \citet{lyu_analytical_2023},  analytical Green's functions can be found for any piecewise linear serrations. Therefore, although the results in this paper are specifically for sawtooth-shaped serrations, similar equations for any piecewise linear serration shapes can be straightforwardly derived in a similar manner. We expect those formulations to have similar computational efficiency as the present model.
	
	\section{Conclusion}
    
	In this work, a three-dimensional semi-analytical model for predicting serrated trailing-edge noise is developed based on the Wiener–Hopf method.
    The recently developed Green’s function~\citep{lyu_analytical_2023}  is used to first calculate the scattered surface pressure over a semi-infinite flat plate with a serrated trailing edge. Far-field sound is obtained by integrating the scattering pressure over a flat plate of finite size. The resulting prediction takes the form of a compact double infinite summation, offering a computationally efficient framework for noise prediction.

    Validation is performed through an examination of the scattered surface pressure distribution and a comparison of noise spectra and directivity prediction with an existing two-dimensional Wiener–Hopf-based model. The scattered-pressure field exhibits a behaviour consistent with expectations along the serrated edge. Comparisons of far-field spectral predictions show close agreement between the three-dimensional and two-dimensional models at moderate observer distances (around $r/c=1$). The three-dimensional formulation predicts systematically lower noise levels at low frequencies, while at higher frequencies it agrees well with two-dimensional predictions. 
    
    Directivity patterns predicted by the two models are also compared, which further illustrates the importance of three-dimensional effects. The present model captures the rapid upstream attenuation and hence a dipolar behaviour expected for finite plates—a feature absent from the two-dimensional formulation. At higher frequencies, both models display broadly similar trends, though the three-dimensional formulation generates additional lobes due to interference patterns induced by the finite span and chord. These differences become more pronounced for sharper serration geometries, highlighting the need for three-dimensional modelling when assessing practical serration designs.

    The computational cost of the present model is two orders of magnitude lower than that of iterative Schwarzschild-based models, rendering it suitable for large-scale parameter sweeps and optimisations. The analytical Green’s-function framework can also be extended to any piecewise linear serration geometry, suggesting wider applicability in future design-oriented works. However, we note that the Green’s function employed here relies on an approximation introduced in the kernel decomposition of \citet{ayton_analytic_2018} and \citet{lyu_analytical_2023}. Consequently, while the present formulation is significantly faster, its accuracy does not consistently match that of the iterative Schwarzschild-based model~\citep{lyu_prediction_2016}. Nevertheless, if optimisation tasks depend primarily on capturing the correct parametric trends of serration geometries, the present model may be a useful tool to balance between accuracy and efficiency. In contrast, the Schwarzschild-based model is likely preferable if a more accurate noise prediction, instead of computational efficiency, is desired. 

\section*{Code Availability}
The source code for this project is available at: \url{https://github.com/sichengZhang-WPG/Sawtooth-TEnoise-prediction.git}

\begin{acknowledgments}
The authors wish to gratefully acknowledge the National Natural Science Foundation of China (NSFC) under the grant numbers 12472263 and U2570222. The second author (BL) wishes to acknowledge the funding from the Beijing Natural Science Foundation (L253027) and from Laoshan Laboratory (LSKJ202202000).
\end{acknowledgments}

\appendix

\section{The $n=0$ result for the transfer function $p_f$}
\label{appendix:n0}
        When $n=0$, we define the coefficients as
	\begin{IEEEeqnarray}{rCl}
            \sigma_{0} &=& \dfrac{k_1}{\beta}+\dfrac{kx_1}{\beta^2S_0},\IEEEnonumber*\\
            \Omega_0 &=& -\chi_0+k\dfrac{x_2}{S_0},\\
            \gamma_{0} &=& \dfrac{\kappa_0-k_1}{\beta},\\
            D_{01} &=& \dfrac{1}{S_0^2}\uexp\bigg\{-\ui k\dfrac{M_0x_1-S_0}{\beta^2}\bigg\}\dfrac{\ui+1}{4(\kappa_0-k_1)\uph},\IEEEyesnumber\\
            D_{02} &=& \dfrac{1}{S_0^2}\uexp\bigg\{-\ui k\dfrac{M_0x_1-S_0}{\beta^2}\bigg\}\dfrac{1-\ui}{2h},\\
            D_{03} &=& \dfrac{1}{S_0^2}\uexp\bigg\{-\ui k\dfrac{M_0x_1-S_0}{\beta^2}\bigg\}\dfrac{\ui-1}{2},\\
            D_{04} &=& -\dfrac{1}{S_0^2}\uexp\bigg\{-\ui k\dfrac{M_0x_1-S_0}{\beta^2}\bigg\}\dfrac{1+\ui}{2h}.\\
    \end{IEEEeqnarray}
    Subsequently, the $n=0$ terms of equation~\eqref{eqn:pf_pre1} can be written as 
	\begin{IEEEeqnarray*}{rCl}
            p_f^{(0,1)}&=\dfrac{\ui\omega x_3}{4\uppi c_0}\sum_{m=-N}^{m=N}D_{01}\bigg\{&\int_{m-\frac{1}{4}}^{m+\frac{3}{4}}\ue^{-\ui \Omega_0y_2}\int_{-c}^{hF(y_2)}\ue^{-\ui \sigma_{0}y_1}E(\gamma_{0}(-y_1+h))\ud y_1\ud y_2\\
            &&\<-\int_{m-\frac{1}{4}}^{m+\frac{3}{4}}\ue^{-\ui \Omega_0y_2}\int_{-c}^{-h}\ue^{-\ui \sigma_{0}y_1}E(\gamma_{0}(-y_1-h))\ud y_1\ud y_2\bigg\},\label{eqn:pf01}\IEEEyesnumber\\
            p_f^{(0,2)}&=\dfrac{\ui\omega x_3}{4\uppi c_0}\sum_{m=-N}^{m=N}D_{02}\bigg\{&\int_{m-\frac{1}{4}}^{m+\frac{3}{4}}\ue^{-\ui \Omega_0y_2}\int_{-c}^{hF(y_2)}y_1\ue^{-\ui \sigma_{0}y_1}E(\gamma_{0}(-y_1+h))\ud y_1\ud y_2\\
            &&\<-\int_{m-\frac{1}{4}}^{m+\frac{3}{4}}\ue^{-\ui \Omega_0y_2}\int_{-c}^{-h}y_1\ue^{-\ui \sigma_{0}y_1}E(\gamma_{0}(-y_1-h))\ud y_1\ud y_2\bigg\},\label{eqn:pf02}\IEEEyesnumber\\
            p_f^{(0,3)}&=\dfrac{\ui\omega x_3}{4\uppi c_0}\sum_{m=-N}^{m=N}D_{03}\bigg\{&\int_{m-\frac{1}{4}}^{m+\frac{3}{4}}\ue^{-\ui \Omega_0y_2}\int_{-c}^{hF(y_2)}\ue^{-\ui \sigma_{0}y_1}E(\gamma_{0}(-y_1+h))\ud y_1\ud y_2\\
            &&\<+\int_{m-\frac{1}{4}}^{m+\frac{3}{4}}\ue^{-\ui \Omega_0y_2}\int_{-c}^{-h}\ue^{-\ui \sigma_{0}y_1}E(\gamma_{0}(-y_1-h))\ud y_1\ud y_2\bigg\},\label{eqn:pf03}\IEEEyesnumber\\
            p_f^{(0,4)}&=\dfrac{\ui\omega x_3}{4\uppi c_0}\sum_{m=-N}^{m=N}D_{04}\bigg\{&\int_{m-\frac{1}{4}}^{m+\frac{3}{4}}\ue^{-\ui \Omega_0y_2}\int_{-c}^{hF(y_2)}\sqrt{\dfrac{-y_1+h}{2\uppi \gamma_0}}\ue^{\ui(\gamma_{0}(-y_1+h))}\ue^{-\ui \sigma_{0}y_1}\ud y_1\ud y_2\\
            &&\<-\int_{m-\frac{1}{4}}^{m+\frac{3}{4}}\ue^{-\ui \Omega_0y_2}\int_{-c}^{-h}\sqrt{\dfrac{-y_1-h}{2\uppi \gamma_0}}\ue^{\ui(\gamma_{0}(-y_1-h))}\ue^{-\ui \sigma_{0}y_1}\ud y_1\ud y_2\bigg\}.\label{eqn:pf04}\IEEEyesnumber\\        
    \end{IEEEeqnarray*}
    
    Now we need to analytically derive the result of the above integral \eqref{eqn:pf01} - \eqref{eqn:pf04}. The form of equation~\eqref{eqn:pf01} is the same as Equation~\eqref{eqn:pf_n}, which leads to a similar result, i.e.
    \begin{IEEEeqnarray}{rCl}
        &&p_f^{(0,1)} = \dfrac{\sin(\Omega_0(N+\frac{1}{2}))}{\sin{(\Omega_0/2)}}\dfrac{\ui\omega x_3}{4\uppi c_0}g^{(0,1)},\label{eqn:g01}\\
        &&p_f^{(0,2)} = \dfrac{\sin(\Omega_0(N+\frac{1}{2}))}{\sin{(\Omega_0/2)}}\dfrac{\ui\omega x_3}{4\uppi c_0}g^{(0,2)},\label{eqn:g02}\\
        &&p_f^{(0,3)} = \dfrac{\sin(\Omega_0(N+\frac{1}{2}))}{\sin{(\Omega_0/2)}}\dfrac{\ui\omega x_3}{4\uppi c_0}g^{(0,3)},\label{eqn:g03}\\
        &&p_f^{(0,4)} = \dfrac{\sin(\Omega_0(N+\frac{1}{2}))}{\sin{(\Omega_0/2)}}\dfrac{\ui\omega x_3}{4\uppi c_0}g^{(0,4)}.\label{eqn:g04}\\
    \end{IEEEeqnarray}
    where
    \begin{IEEEeqnarray*}{rCl}
            g^{(0,1)}&=&
            \dfrac{4D_{01}h\ue^{-\ui \Omega_0/4}\ue^{-\ui \sigma_0h}}{\Omega_0(\Omega_0\pm4\sigma_0h)}\Q\left(4\gamma_0,(4\sigma_0\pm \Omega_0/h),h\right)\\
            &&\<\pm\dfrac{D_{01}\ue^{-\ui \Omega_0/4}\ue^{\pm\ui \frac{\Omega_0}{2}}}{\sigma_0}\left\{\dfrac{\ue^{\ui \sigma_0h}}{\Omega_0\pm4\sigma_0h}\Q\left(4\gamma_0,0,h\right)-\dfrac{\ue^{-\ui \sigma_0h}}{\Omega_0}\Q(4\gamma_0,4\sigma_0,h)\right\}\IEEEyesnumber\label{eqn:pf01_int}\\
            &&\<\pm\dfrac{\ui D_{01}\ue^{-\ui \Omega_0/4}}{\sigma_0}\dfrac{2}{\Omega_0}\sin{\dfrac{\Omega_0}{2}}\bigg\{\ue^{\mp\ui \sigma_0h}\Q\left(2\gamma_0,2\sigma_0,c\pm h\right)-\ue^{\ui \sigma_0c}\Q\left(2\gamma_0,0,c\pm h\right)\bigg\}.
    \end{IEEEeqnarray*}

    The distinction between the forms of equation~\eqref{eqn:pf01} and \eqref{eqn:pf03} lies only in the sign of the second double integral. Using the same method of integration by parts, or just altering the sign of the terms in equation~\eqref{eqn:pf01_int} with the coefficient $(c-h)$, $g^{(0,3)}$in equation~\eqref{eqn:g03} can be evaluated to give
    \begin{IEEEeqnarray*}{rCl}
            g^{(0,3)}&=&\dfrac{4D_{03}h\ue^{-\ui \Omega_0\frac{1}{4}}\ue^{-\ui \sigma_0h}}{\Omega_0(\Omega_0\pm4\sigma_0h)}\Q\left(4\gamma_0,(4\sigma_0\pm \Omega_0/h),h\right)\\
            &&\<\pm\dfrac{D_{03}\ue^{-\ui \Omega_0\frac{1}{4}}\ue^{\pm\ui \frac{\Omega_0}{2}}}{\sigma_0}\left\{\dfrac{\ue^{\ui \sigma_0h}}{\Omega_0\pm4\sigma_0h}\Q\left(4\gamma_0,0,h\right)-\dfrac{\ue^{-\ui \sigma_0h}}{\Omega_0}\Q(4\gamma_0,4\sigma_0,h)\right\}\IEEEyesnumber\label{eqn:pf03_int}\\
            &&\<+\dfrac{\ui D_{03}\ue^{-\ui \Omega_0\frac{1}{4}}}{\sigma_0}\dfrac{2}{\Omega_0}\sin{\dfrac{\Omega_0}{2}}\left\{\ue^{\mp\ui \sigma_0h}\Q\left(2\gamma_0,2\sigma_0,c\pm h\right)-\ue^{\ui \sigma_0c}\Q\left(2\gamma_0,0,c\pm h\right)\right\}.
    \end{IEEEeqnarray*}
    
    Equation~\eqref{eqn:pf02} can be assessed by computing the partial derivative of equation~\eqref{eqn:pf01} with respect to $\sigma_0$, i.e. 
    \begin{IEEEeqnarray*}{rCl}
            p_f^{(0,2)}&=&\ui \dfrac{D_{02}}{D_{01}}\dfrac{\partial}{\partial \sigma_0}p_f^{(0,1)}.
    \end{IEEEeqnarray*}
    Then $g^{(0,2)}$ in equation~\eqref{eqn:g02} can be derived as
    \begin{IEEEeqnarray*}{rCl}
            g^{(0,2)}&=&D_{02}\dfrac{2}{\Omega_0}\sin{\dfrac{\Omega_0}{2}}\ue^{-\ui \Omega_0/4}\bigg\{\left[\dfrac{\ui h}{\sigma_0}\pm\dfrac{1}{\sigma_0^2}\right]\ue^{\mp\ui \sigma_0h}\Q(2\gamma_0,2\sigma_0,c\pm h)\\
            &&\<\pm\left[\dfrac{\ui c}{\sigma_0}-\dfrac{1}{\sigma_0^2}\right]\ue^{\ui \sigma_0c}\Q(2\gamma_0,0,c\pm h) \pm\dfrac{1}{2\sigma_0}\dfrac{\ue^{\mp\ui \sigma_0h}}{\sigma_0+\gamma_0}\Q(2\gamma_0,2\sigma_0,c\pm h)\\
            &&\<\mp\dfrac{1}{\sigma_0}\sqrt{\dfrac{\gamma_0}{\sigma_0+\gamma_0}}\dfrac{(c\pm h)\ue^{\ui \gamma_0(c\pm h)}\ue^{\ui \sigma_0c}}{\sqrt{2\uppi(\sigma_0+\gamma_0)(c\pm h)}}\bigg\}\\
            &&\<+\ui D_{02}\ue^{-\ui \Omega_0/4}\dfrac{\ue^{\ui \sigma_0h}}{\sigma_0}\dfrac{\ue^{\pm\ui\frac{\Omega_0}{2}}}{\Omega_0\pm4\sigma_0h}\left[\pm\ui h\mp\dfrac{1}{\sigma_0}-\dfrac{4h}{\Omega_0\pm4\sigma_0h}\right]\Q(4\gamma_0,0,h)\IEEEeqnarraynumspace\\
            &&\<+\frac{\ui D_{02}\ue^{-\ui \Omega_0/4}\ue^{-\ui \sigma_0h}}{\sigma_0(\Omega_0\pm 4\sigma_0h)}\left[\pm\ui h\pm\frac{1}{\sigma_0} +\dfrac{4h}{\Omega_0\pm4\sigma_0h}\right]\Q\left(4\gamma_0,4\sigma_0\pm\Omega_0/h,h\right)\\
            &&\<\pm\frac{\ui D_{02}\ue^{-\ui \Omega_0/4}2h\ue^{-\ui \sigma_0h}}{\sigma_0(\Omega_0\pm 4\sigma_0h)(4\sigma_0h+4\gamma_0h\pm \Omega_0)}\Q\left(4\gamma_0,4\sigma_0\pm\Omega_0/h,h\right)\IEEEyesnumber\\
            &&\<\mp\frac{\ui D_{02}\ue^{-\ui \Omega_0/4}2h\sqrt{4\gamma_0h}\ue^{-\ui \sigma_0h}}{\sigma_0(\Omega_0\pm 4\sigma_0h)\sqrt{4\sigma_0h+4\gamma_0h\pm \Omega_0}}\dfrac{\ue^{\frac{\ui}{2}(4\sigma_0h+4\gamma_0h\pm \Omega_0)}}{\sqrt{\uppi(4\sigma_0h+4\gamma_0h\pm \Omega_0)}}\\
            &&\<\mp\frac{\ui D_{02}\ue^{-\ui \Omega_0/4}(\ui h +1/\sigma_0)\ue^{-\ui \sigma_0h}}{\sigma_0\Omega_0}\Q\left(4\gamma_0,4\sigma_0\pm\Omega_0/h,h\right)\\
            &&\<\mp\frac{\ui D_{02}\ue^{-\ui \Omega_0/4}2h\ue^{-\ui \sigma_0h}}{\sigma_0\Omega_0(4\sigma_0h+4\gamma_0h\pm \Omega_0)}\Q\left(4\gamma_0,4\sigma_0\pm\Omega_0/h,h\right)\\
            &&\<\pm\frac{\ui D_{02}\ue^{-\ui \Omega_0/4}2h\sqrt{4\gamma_0h}\ue^{-\ui \sigma_0h}}{\sigma_0\Omega_0\sqrt{4\sigma_0h+4\gamma_0h\pm \Omega_0}}\dfrac{\ue^{\frac{\ui}{2}(4\sigma_0h+4\gamma_0h\pm \Omega_0)}}{\sqrt{\uppi(4\sigma_0h+4\gamma_0h\pm \Omega_0)}}\\ 
            &&\<+\frac{D_{02}}{\Omega_0}\ue^{-\ui \Omega_0/4}\sin{\dfrac{\Omega_n}{2}}\Bigg\{\left[-\ui h-\frac{1}{\sigma_0}-\dfrac{1}{2(\sigma_0+\gamma_0)}\right]\dfrac{\ue^{-\ui \sigma_0h}}{\sigma_0}\Q(4\gamma_0,4\sigma_0,h)\\
            &&\<+\dfrac{\ue^{-\ui \sigma_0h}}{\sigma_0}\dfrac{2h\sqrt{\gamma_0}}{\sqrt{\sigma_0+\gamma_0}}\dfrac{\ue^{2\ui(\sigma_0+\gamma_0)h}}{\sqrt{4\uppi(\sigma_0+\gamma_0)h}}\Bigg\}.
    \end{IEEEeqnarray*}
    
    Equation~\eqref{eqn:pf04} can be evaluated by taking the partial derivative of equation~\eqref{eqn:pf01} with respect to $\gamma_0$, i.e. 
    \begin{IEEEeqnarray*}{rCl}
            p_f^{(0,4)}&=&\dfrac{D_{04}}{D_{01}}\dfrac{\partial}{\partial \gamma_0}p_f^{(0,1)}.
    \end{IEEEeqnarray*}
    Then $g^{(0,4)}$ in equation~\eqref{eqn:g04} can be derived as
    \begin{IEEEeqnarray*}{rCl}
            g^{(0,4)}&=&D_{04}\dfrac{2}{\Omega_0}\sin{\dfrac{\Omega_0}{2}}\ue^{-\ui \Omega_0/4}\dfrac{\ui}{\sigma_0}\bigg\{\pm\dfrac{\sigma_0\ue^{\mp\ui \sigma_0h}}{2\gamma_0(\sigma_0+\gamma_0)}\Q(2\gamma_0,2\sigma_0,c\pm h)\\
            &&\<\pm\sqrt{\dfrac{\gamma_0}{\sigma_0+\gamma_0}}\ue^{\mp\ui \sigma_0h}\dfrac{(c\pm h)\ue^{\ui(\sigma_0+\gamma_0)(c\pm h)}}{\sqrt{2\uppi(\sigma_0+\gamma_0)(c\pm h)}} \mp \ue^{\ui \sigma_0c}\dfrac{(c\pm h)\ue^{\ui \gamma_0(c\pm h)}}{\sqrt{2\uppi \gamma_0(c\pm h)}}\bigg\}\\
            &&\<\pm\dfrac{D_{04}\ue^{-\ui \Omega_0/4}\ue^{\ui \sigma_0h}2h\ue^{\ui2\gamma_0h}}{\sigma_0\sqrt{4\uppi\gamma_0h}}\dfrac{\ue^{\pm\ui\frac{\Omega_0}{2}}}{\Omega_0\pm4\sigma_0h}\\
            &&\<\pm\dfrac{D_{04}\ue^{-\ui \Omega_0/4}\ue^{-\ui \sigma_0h}2h}{\Omega_0\gamma_0(4\sigma_0h+4\gamma_0h\pm \Omega_0)}\Q\left(4\gamma_0,4\sigma_0\pm\Omega_0/h,h\right)\IEEEyesnumber\\
            &&\<\mp\frac{D_{04}\ue^{-\ui \Omega_0/4}2h\sqrt{4\gamma_0h}\ue^{-\ui \sigma_0h}}{\sigma_0(\Omega_0\pm4\sigma_0h)\sqrt{4\sigma_0h+4\gamma_0h\pm \Omega_0}}\dfrac{\ue^{\frac{\ui}{2}(4\sigma_0h+4\gamma_0h\pm \Omega_0)}}{\sqrt{\uppi(4\sigma_0h+4\gamma_0h\pm \Omega_0)}}\\
            &&\<\pm\frac{D_{04}\ue^{-\ui \Omega_0/4}2h\sqrt{4\gamma_0h}\ue^{-\ui \sigma_0h}}{\sigma_0\Omega_0\sqrt{4\sigma_0h+4\gamma_0h\pm \Omega_0}}\dfrac{\ue^{\frac{\ui}{2}(4\sigma_0h+4\gamma_0h\pm \Omega_0)}}{\sqrt{\uppi(4\sigma_0h+4\gamma_0h\pm \Omega_0)}}\\
            &&\<\mp\dfrac{D_{04}\ue^{-\ui \Omega_0/4}\ue^{-\ui \sigma_0h}\ue^{\pm\ui\frac{\Omega_0}{2}}}{\sigma_0\Omega_0}\left[\dfrac{\sigma_0\Q(4\gamma_0,4\sigma_0,h)}{2\gamma_0(\sigma_0+\gamma_0)}+\sqrt{\dfrac{\gamma_0}{\sigma_0+\gamma_0}}\dfrac{2h\ue^{\ui2(\sigma_0+\gamma_0)h}}{\sqrt{4\uppi(\sigma_0+\gamma_0)h}}\right].\IEEEeqnarraynumspace
    \end{IEEEeqnarray*}

\bibliography{apssamp}

@inproceedings{li_extensions_2022,
	address = {Southampton, UK},
	title = {Extensions and {Applications} of {Lyu} and {Ayton}'s {Serrated} {Trailing}-{Edge} {Noise} {Model} to {Rotorcraft}},
	isbn = {978-1-62410-664-4},
	booktitle = {28th {AIAA}/{CEAS} {Aeroacoustics} 2022 {Conference}},
	publisher = {American Institute of Aeronautics and Astronautics},
	author = {Li, Sicheng (Kevin) and Lee, Seongkyu},
	month = jun,
	year = {2022},
}

@article{ZHOU2025110851,
title = {An experimental investigation into the aerofoil trailing-edge broadband noise reduction using ogee serrations},
journal = {Applied Acoustics},
volume = {239},
pages = {110851},
year = {2025},
issn = {0003-682X},
author = {Wenhao Zhou and Fusheng Qiu and Yulong Sun and Tongyang Shi and Qingqing Ye and Benshuai Lyu},
}

@article{Tian_Lyu_2024, title={The impact of non-frozen turbulence on the modelling of the noise from serrated trailing edges}, volume={990}, journal={Journal of Fluid Mechanics}, author={Tian, H. and Lyu, B.}, year={2024}, pages={A4}}

@article{arce_leon_flow_2016,
	title = {Flow topology and acoustic emissions of trailing edge serrations at incidence},
	volume = {57},
	issn = {0723-4864, 1432-1114},
	language = {english},
	number = {5},
	urldate = {2025-11-08},
	journal = {Experiments in Fluids},
	author = {Arce León, Carlos and Ragni, Daniele and Pröbsting, Stefan and Scarano, Fulvio and Madsen, Jesper},
	month = may,
	year = {2016},
	pages = {91},
}

@inbook{noble_methods_1959,
	address = {London},
	title = {Methods based on the {Wiener}‐{Hopf} technique for the solution of partial differential equations},
	publisher = {Pergamon Press},
	author = {Noble, Ben and Weiss, George},
	year = {1959},
}

@article{tian2022theoretical,
	title={Theoretical investigation of noise from rotating blades with serrated trailing edges},
	author={Tian, Haopeng and Lyu, Benshuai},
	journal={28th AIAA/CEAS Aeroacoustics 2022 Conference},
	year = {2022},
	pages = {3091},
}

@article{lyu_prediction_2016,
	title = {Prediction of noise from serrated trailing edges},
	volume = {793},
	issn = {1469-7645},
	journal = {Journal of Fluid Mechanics},
	author = {Lyu, B. and Azarpeyvand, M. and Sinayoko, S.},
	year = {2016},
	pages = {556--588},
}

@article{roger_back-scattering_2005,
	title = {Back-scattering correction and further extensions of {Amiet}'s trailing-edge noise model. {Part} 1: theory},
	volume = {286},
	issn = {0022-460X},
	number = {3},
	journal = {Journal of Sound and Vibration},
	author = {Roger, Michel and Moreau, Stéphane},
	month = sep,
	year = {2005},
	pages = {477--506},
}

@article{amiet_noise_1976,
	title = {Noise due to turbulent flow past a trailing edge},
	volume = {47},
	issn = {0022-460X},
	number = {3},
	journal = {Journal of sound and vibration},
	author = {Amiet, Roy K},
	year = {1976},
	pages = {387--393},
}

@article{amiet_effect_1978,
	title = {Effect of the incident surface pressure field on noise due to turbulent flow past a trailing edge.},
	author = {Amiet, RK},
	year = {1978},
	volume = {57},
	pages = {305--306},
	journal = {Journal of Sound and Vibration},
}

@article{lyu_analytical_2023,
	title = {Analytical {Green}'s function for the acoustic scattering by a flat plate with a serrated edge},
	volume = {961},
	issn = {1469-7645},
	journal = {Journal of Fluid Mechanics},
	author = {Lyu, B.},
	year = {2023},
	pages = {A33},
}

@article{ayton_analytic_2018,
	title = {Analytic solution for aerodynamic noise generated by plates with spanwise-varying trailing edges},
	volume = {849},
	issn = {1469-7645},
	journal = {Journal of Fluid Mechanics},
	author = {Ayton, Lorna J.},
	year = {2018},
	pages = {448--466},
}

@article{chase1987character,
	title={The character of the turbulent wall pressure spectrum at subconvective wavenumbers and a suggested comprehensive model},
	author={Chase, DM},
	journal={Journal of Sound and Vibration},
	volume={112},
	number={1},
	pages={125--147},
	year={1987},
	publisher={Elsevier}
}

@article{tian_prediction_2022,
	title = {Prediction of broadband noise from rotating blade elements with serrated trailing edges},
	volume = {34},
	issn = {1070-6631},
	number = {8},
	urldate = {2023-11-30},
	journal = {Physics of Fluids},
	author={Tian, Haopeng and Lyu, Benshuai},
	month = aug,
	year = {2022},
	pages = {085109},
}

@article{avallone_benefits_2017,
	title = {Benefits of curved serrations on broadband trailing-edge noise reduction},
	volume = {400},
	issn = {0022-460X},
	journal = {Journal of Sound and Vibration},
	author = {Avallone, F and Van der Velden, WCP and Ragni, D},
	year = {2017},
	pages = {167--177},
}

@article{avallone_noise_2018,
	title = {Noise reduction mechanisms of sawtooth and combed-sawtooth trailing-edge serrations},
	volume = {848},
	issn = {0022-1120},
	journal = {Journal of Fluid Mechanics},
	author = {Avallone, F and Van Der Velden, WCP and Ragni, D and Casalino, D},
	year = {2018},
	pages = {560--591},
}

@article{chong_airfoil_2013,
	title = {Airfoil self noise reduction by non-flat plate type trailing edge serrations},
	volume = {74},
	issn = {0003-682X},
	number = {4},
	journal = {Applied Acoustics},
	author = {Chong, TP and Joseph, PF and Gruber, M},
	year = {2013},
	pages = {607--613},
}

@article{chong_aeroacoustic_2015,
	title = {On the aeroacoustic and flow structures developed on a flat plate with a serrated sawtooth trailing edge},
	volume = {354},
	issn = {0022-460X},
	journal = {Journal of Sound and Vibration},
	author = {Chong, Tze Pei and Vathylakis, Alexandros},
	year = {2015},
	pages = {65--90},
}

@article{dassen_results_1996,
	title = {Results of a wind tunnel study on the reduction of airfoil self-noise by the application of serrated blade trailing edges},
	journal = {Proceeding of the European Union Wind Energy Conference and Exhibition},
	author = {Dassen, T and Parchen, R and Bruggeman, J and Hagg, F},
	year = {1996},
	pages = {800-803}
}

@article{gruber_airfoil_2012,
	title = {Airfoil noise reduction by edge treatments},
	author = {Gruber, Mathieu},
	journal = {PhD thesis, University of Southampton, University Road, Southampton SO17, 1BJ},
	year = {2012},
}

@article{gruber_airfoil_2010,
	title = {Airfoil trailing edge noise reduction by the introduction of sawtooth and slitted trailing edge geometries},
	volume = {10},
	journal = {Proceedings of 20th International Congress on Acoustics},
	author = {Gruber, Mathieu and Azarpeyvand, Mahdi and Joseph, Phillip F},
	year = {2010},
	pages = {6},
}

@article{howe_review_1978,
	title = {A review of the theory of trailing edge noise},
	volume = {61},
	issn = {0022-460X},
	number = {3},
	journal = {Journal of sound and vibration},
	author = {Howe, Michael S},
	year = {1978},
	pages = {437--465},
}

@article{howe_aerodynamic_1991,
	title = {Aerodynamic noise of a serrated trailing edge},
	volume = {5},
	issn = {0889-9746},
	number = {1},
	journal = {Journal of Fluids and Structures},
	author = {Howe, Michael S},
	year = {1991},
	pages = {33--45},
}

@article{howe_noise_1991,
	title = {Noise produced by a sawtooth trailing edge},
	volume = {90},
	issn = {0001-4966},
	number = {1},
	journal = {The Journal of the Acoustical society of America},
	author = {Howe, Michael S},
	year = {1991},
	pages = {482--487},
}

@article{jaworski_aeroacoustics_2020,
	title = {Aeroacoustics of silent owl flight},
	volume = {52},
	issn = {0066-4189},
	journal = {Annual Review of Fluid Mechanics},
	author = {Jaworski, Justin W and Peake, Nigel},
	year = {2020},
	pages = {395--420},
}

@article{jones_acoustic_2012,
	title = {Acoustic and hydrodynamic analysis of the flow around an aerofoil with trailing-edge serrations},
	volume = {706},
	issn = {1469-7645},
	journal = {Journal of Fluid Mechanics},
	author = {Jones, LE and Sandberg, RD},
	year = {2012},
	pages = {295--322},
}

@article{leon_effect_2017,
	title = {Effect of trailing edge serration-flow misalignment on airfoil noise emissions},
	volume = {405},
    language = {english},
	issn = {0022-460X},
	journal = {Journal of Sound and Vibration},
	author = {León, Carlos Arce and Merino-Martínez, Roberto and Ragni, Daniele and Avallone, Francesco and Scarano, Fulvio and Pröbsting, Stefan and Snellen, Mirjam and Simons, Dick G and Madsen, Jesper},
	year = {2017},
	pages = {19--33},
}

@article{lyu_rapid_2020,
	title = {Rapid noise prediction models for serrated leading and trailing edges},
	volume = {469},
	issn = {0022-460X},
	journal = {Journal of Sound and Vibration},
	author = {Lyu, Benshuai and Ayton, Lorna J},
	year = {2020},
	pages = {115--136},
}

@article{moreau_noise-reduction_2013,
	title = {Noise-reduction mechanism of a flat-plate serrated trailing edge},
	volume = {51},
	issn = {0001-1452},
	number = {10},
	journal = {AIAA journal},
	author = {Moreau, Danielle J and Doolan, Con J},
	year = {2013},
	pages = {2513--2522},
}

@article{oerlemans_reduction_2009,
	title = {Reduction of wind turbine noise using optimized airfoils and trailing-edge serrations},
	volume = {47},
	issn = {0001-1452},
	number = {6},
	journal = {AIAA journal},
	author = {Oerlemans, Stefan and Fisher, Murray and Maeder, Thierry and Kögler, Klaus},
	year = {2009},
	pages = {1470--1481},
}

@inproceedings{sanjose_direct_2014,
	title = {Direct numerical simulation of acoustic reduction using serrated trailing-edge on an isolated airfoil},
	booktitle = {Proceedings of 20th AIAA/CEAS Aeroacoustics Conference},
	author = {Sanjosé, Marlène and Méon, Claire and Moreau, Stephane and Idier, Alexandre and Laffay, Paul},
	year = {2014},
	pages = {2324},
}

\end{document}